\documentclass[11pt]{article}
\usepackage{wrapfig}
\usepackage[centertags]{amsmath}
\usepackage{amssymb}
\usepackage[pdftex,dvipsnames]{xcolor}
\usepackage{graphicx}
\usepackage[us]{datetime}
\usepackage{afterpage}
\usepackage{enumitem}
\usepackage[normalem]{ulem}
\usepackage[Symbol]{upgreek}
\usepackage{bm}
\usepackage[export]{adjustbox}
\usepackage{multirow}
\usepackage{comment}
\usepackage{float}
\usepackage[psamsfonts,mathscr]{eucal}
\usepackage[]{graphicx}
\usepackage{color}
\usepackage{enumitem}
\usepackage[utf8]{inputenc}
\usepackage[titletoc]{appendix}
\usepackage{cases}

\usepackage{algorithmicx}
\usepackage{algorithm}
\usepackage{algcompatible}

\graphicspath{{Figures/}}
\usepackage{epsfig}
\usepackage[margin=1.25in]{geometry}
\usepackage[labelfont=bf]{caption}
\usepackage{setspace}

\usepackage[labelformat=parens]{subcaption}
\usepackage{booktabs} 

\definecolor{PlainBlue}{rgb}{0,0,.7}
\definecolor{PlainRed}{rgb}{.7,0,0}
\usepackage[colorlinks=true,citecolor=ForestGreen,linkcolor=PlainRed,urlcolor=PlainBlue,bookmarks=false]{hyperref}
\usepackage[capitalise]{cleveref}

\usepackage{soul}

\newcommand{\mat}[1]{\mathsf{#1}}

\usepackage[colorinlistoftodos,prependcaption,textsize=scriptsize,textwidth=1.05in]{todonotes}
\usepackage{xargs} 
\newcommandx{\jpc}[2][1=]{\todo[linecolor=OliveGreen,bordercolor=OliveGreen,
backgroundcolor=OliveGreen!10,caption=\textbf{JPC},#1]{#2}}

\usepackage[numbers,sort&compress]{natbib}

\title{\bf Model Reduction of a Flexible Nonsmooth Oscillator Recovers its Entire\\
Bifurcation Structure}

\author{Suparno Bhattacharyya and Joseph.~P.~Cusumano\thanks{Address correspondence to:~\href{mailto:jpc3@psu.edu}{\tt jpc3@psu.edu}.}\\[8pt]
Department of Engineering Science and Mechanics\\
Pennsylvania State University}

\date{June 19, 2025 \\[4pt]}

\begin{document}

\maketitle

\vspace{-1\baselineskip}
\begin{abstract}
We study the reduced order modeling of a piecewise-linear, globally nonlinear flexible oscillator in which a Bernoulli-Euler beam is subjected to a position-triggered kick force and a piecewise restoring force at its tip. The nonsmooth boundary conditions, which determine different regions of a hybrid phase space, can generally be expected to excite many degrees of freedom. With kick strength as parameter, the system's bifurcation diagram is found to exhibit a range of periodic and chaotic behaviors. Proper orthogonal decomposition (POD) is used to
obtain a single set of global basis functions spanning all of the hybrid regions. The reduced order model (ROM) dimension is chosen using previously developed energy closure analysis, ensuring approximate energy balance on the reduced subspace. This yields accurate ROMs with 8 degrees of freedom. Remarkably, we find that ROMs formulated using using data from individual periodic steady states can nevertheless be used to reconstruct the entire bifurcation structure of the original system without updating. This demonstrates that, despite being constructed with steady state data, the ROMs model sufficiently small transients with enough accuracy to permit using simple continuation for the bifurcation diagram. We also find ROM subspaces obtained for different values of the bifurcation parameter are essentially identical. Thus, POD augmented with energy closure analysis is found to reliably yield effective dimension estimates and ROMs for this nonlinear, nonsmooth system that are robust across stability transitions, including even period doubling cascades to chaos, thereby greatly reducing data requirements and computational costs. 
\end{abstract}

{\textbf{keywords}:  dimension reduction, physics-informed surrogate modeling, proper orthogonal decomposition, nonlinear vibrations, nonsmooth oscillators, closure models.}

\section{Introduction}
\label{sec:intro}

Reduced order models (ROMs) approximate the dominant behavior of high-dimensional dynamical systems while preserving their key characteristics. This work investigates the projection-based model order reduction of an infinite-dimensional hybrid dynamical system: a piecewise linear, globally nonlinear flexible mechanical oscillator. 
Computational analysis of such systems is often resource-intensive, as numerical techniques such as the finite element method generally require a substantial number of degrees of freedom (DOFs) to attain high-fidelity solutions. Consequently, accurate ROMs are crucial for efficient analysis. Here, we employ proper orthogonal decomposition (POD) in conjunction with a previously developed energy closure analysis for estimating the degree of reduction, yielding  reduced order models that exhibit high accuracy. Furthermore, we show that a ROM estimated with data from only a single periodic steady state is capable of reconstructing the entire bifurcation structure of the original oscillator. Thus, only a single ROM need be estimated to model the entire range of system behaviors as parameters are varied, including stability transitions. This results in a great savings in data requirements and computational costs needed to create ROMs. 

Projection-based reduced order modeling \cite{benner_survey_2015} is a prominent model reduction approach in which the solution to high-dimensional systems is approximated through a least-squares optimal projection of the governing equations onto a low-dimensional linear subspace that captures the essential dynamics of the system \cite{sirovich_turbulence_1987}. 
Proper orthogonal decomposition (POD) \cite{berkooz_proper_1993, feeny_proper_2002, chelidze_smooth_2006, shlizerman2012proper, sirovich_chaotic_1989, sirovich_turbulence_1987, azeez_proper_2001, Han2003, kerschen_method_2005, Khalil2007, kostova-vassilevska_model_2018, jin_adaptive_2017}, also known as the Karhunen-Loève decomposition \cite{bellizzi_karhunenloeve_2009, bellizzi_smooth_2009, breuer_use_1991, everson1995karhunen, newman_model_1996, newman_model_1996-1, wolter_reduced-order_2002}, is widely regarded as the standard method for identifying such subspaces. 
However, accurately estimating the dimension of these reduced models, a critical factor influencing their accuracy, remains a significant challenge. 
Conventionally, the dimension of a POD-based ROM is determined by selecting a subspace that captures a specified percentage of the variance in the data obtained from the high-dimensional system \cite{sirovich_turbulence_1987, berkooz_proper_1993}. 
In prior work \cite{Bhattacharyya_2020,Bhattacharyya_2022}, we highlighted the limitations of this variance-based dimension estimation by studying a linear structural system subjected to impulsive loading. To address these limitations, we proposed a physics-informed criterion as an enhancement to the conventional POD framework, wherein the model dimension is selected based on balancing the energy input and dissipated within the reduced subspace, rather than relying solely on data variance.

This work extends the application of this \textit{energy closure} approach to nonlinear systems, focusing on the model order reduction of a {hybrid} (piecewise smooth) dynamical system that is infinite dimensional from first principles. We further examine the robustness of ROMs so generated with respect to parameter variations, including across stability transitions.
Such hybrid systems are characterized by both time- and event-driven dynamics \cite{goebel2012hybrid,Bernardo_bif_2008}, with discrete events prompting transitions between different dynamical regimes in different regions of the system's state-space. 
The discrete events may involve abrupt changes in boundary conditions, actuator switching, or state-dependent forces, often resulting in nonsmooth dynamics that present significant challenges for both modeling and computational analysis \cite{sahai2013uncertainty}. 

Nonsmooth dynamics are common in mechanical systems, which frequently contain clearances, impacts, piecewise actuation, or sliding friction \cite{Bernardo2008}. Of current interest are applications exploiting bistable and impacting  elements for vibration energy harvesting  \cite{wang2024_asymmetric_BEH} and nonlinear energy absorption \cite{liu2025_physinf_SINDy_NES, liu2024_sparse_bistable_NES_chain}. Systems with impacts appear in both macro- and microscale applications, including robot locomotion \cite{sun2025_vibroimpact_response, yan2025_capsule_robot_vmises}. The kicked flexible oscillator studied here provides a specific infinite dimensional example of such systems. Thus, the reduction methodology we employ can be expected to apply to a broad class of mechanical systems characterized by nonlinearity, discontinuity, and hybrid dynamics.

\color{black}
The model of the kicked oscillator considered here is similar to one studied experimentally \cite{cusumano_experimental_1996, cusumano_nonlinear_2007}. 
In the experimental setup, a vertical cantilever beam with a permanent magnet at its tip interacts with an electromagnet positioned below it. 
This interaction generates an impulsive force near the vertical equilibrium position whenever the tip passes with sufficient velocity.
Furthermore, the restoring force between the permanent magnet and the kicker magnet core diminishes rapidly as the beam tip moves away from the electromagnet.

Our model consists of a Bernoulli-Euler beam with an end mass. In place of a magnetic interaction, we impose a localized, piecewise-constant tip force. 
The model is autonomous and exhibits various steady-state behaviors, including bifurcations leading to chaos, depending on the kick strength. 
Numerous degrees of freedom (DOFs) are excited during transitions that demarcate regions in the system's hybrid state space, making the system's effective dimensionality event-dependent and challenging to estimate a priori for ROM construction. 

Extensive research has been conducted on piecewise oscillators subjected to periodic external forcing and/or impacts (e.g., \cite{shaw1983periodically, moon1983chaotic, natsiavas1989periodic, nordmark1991nonperiodic, peterka1992transition, cusumano_period-infinity_1993, cusumano_experimental_1994, wiercigroch1994a, blankeship1995steady, Yagasaki2004, Thota2006, Ing2008, Moussi2015, Shui2018, Wang2020}), contributing to a well-established theoretical understanding of their dynamics and bifurcations \cite{Bernardo2008, Ibrahim2009, Makarenkov2012, Leine2013}. 
The application of proper orthogonal decomposition (POD) has been previously used for the reduced-order modeling of vibro-impact and piecewise-linear oscillators \cite{ritto_new_2011, ritto_proper_2012, Jung2012_BAA, Clark2015_NSVCFD}. 
Saito and Epureanu \cite{Saito2011_bilinearModal} proposed a ROM based on bilinear modes for elastic structures with localized piecewise-linearity, which yielded modes closely aligned with dominant POD modes. However, compared to its application to smooth systems, the application of POD to hybrid systems remains relatively underexplored. 

The robustness of ROMs with respect to parametric variations, particularly for systems undergoing stability transitions, remains an open question  \cite{terragni_efficient_2014,Terragni2012,Khamlich2022}. For systems undergoing bifurcations, it is not clear if a single ROM, once obtained, can be expected to capture the entire bifurcation structure of the original system or if, instead, different ROMs might be needed for different dynamical regimes. This question is especially acute for infinite-dimensional hybrid systems, such as the one studied here.

Nonlinear manifold-based methods have been used for dimension reduction of nonlinear dynamical systems  \cite{shaw1983periodically, kerschen2009nonlinear, nayfeh1994on, haller2016nonlinear,Karoui2025,Li2024}, including hybrid systems \cite{Bettini2024}. 
In contrast to POD, which merely identifies a linear subspace containing the dynamics, these approaches construct the system dynamics on nonlinear manifolds embedded within the original high-dimensional space. For a given level of accuracy, these nonlinear methods have the potential to generate reduced-order models (ROMs) with greater dimension reduction than that obtainable with linear subspace methods. To achieve this, such methods often involve nonlinear normal form transformations, invariant manifold computations and, in some cases, require system-specific insights to identify meaningful reduced coordinates. More recently, data-driven approaches \cite{tenenbaum2000a,belkin2003laplacian}, including deep-learning methods \cite{lee2020model,hinton2006reducing} have been developed to determine low-dimensional manifolds directly from data. 

While perhaps not capable of the maximum theoretical reduction possible with these nonlinear methods, POD with energy closure \cite{Bhattacharyya_2022,Bhattacharyya_2020} is  conceptually familiar and computationally simpler. It combines the construction of empirical bases using well-understood covariance statistics with a physics-informed subspace selection criterion, while not requiring detailed knowledge of the governing system's nonlinear phase space structure. The incorporation of energy closure crucially links the model's fidelity to the system's underlying physics---an element absent in purely statistical methods---and, as demonstrated here, enables the construction of highly accurate reduced-order models.

The remainder of the paper is structured as follows: we first develop a model for a flexible kicked-oscillator, followed by a discussion of its steady-state dynamics and bifurcation structure. 
We then carry out the model reduction procedure using POD with energy closure dimension estimation. We conclude by simulating the reduced order model and comparing its behavior to the original system.

\section{Model formulation and piecewise solution}
\label{sec:md}

\begin{figure}[t]
	\centering
        \begin{subfigure}[b]{.2\textwidth}
        \centering
        \includegraphics[width=.85\textwidth]{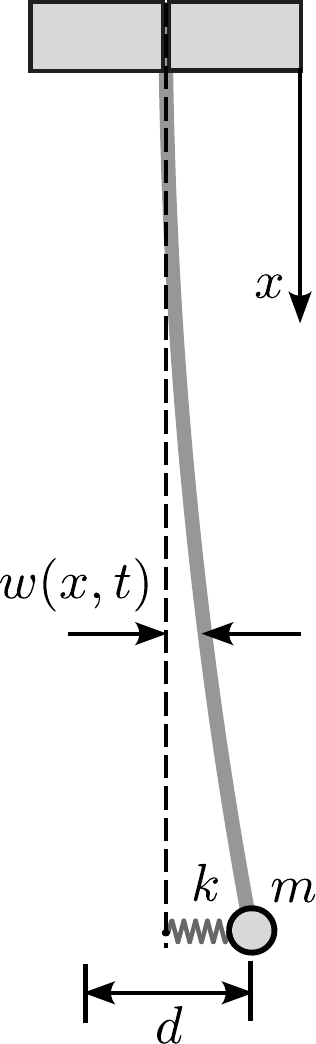}
        \vspace{.5\baselineskip}
        \caption{}
        \label{fig:es_b}
    \end{subfigure}
    \hfill
    \begin{subfigure}[b]{.77\textwidth}
    	\centering
        \includegraphics[width=0.68\textwidth]{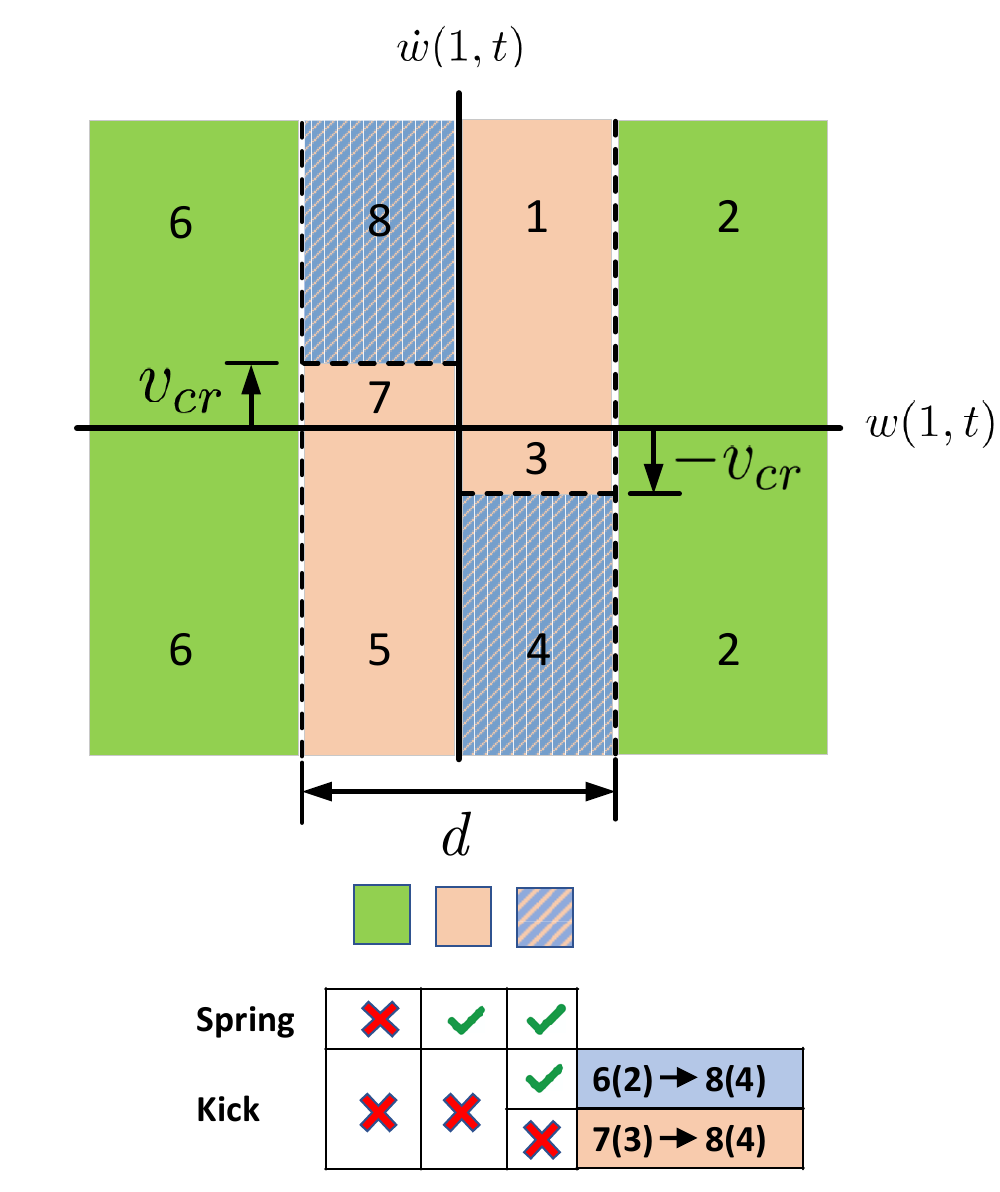}
		\vspace{1\baselineskip}
        \caption{}
        \label{fig:phse}
    \end{subfigure}
    \caption{Model of a vertical kicked beam flexible oscillator: (a) Bernoulli-Euler beam model with displacement \( w(x,t) \). The beam tip has a point mass \( m \) connected to ground by a piecewise spring of stiffness \( k \), with \( d \) representing the size of the region where the kick force and spring are active. These hybrid boundary conditions arise from these piecewise spring and applied forces (Eqs.~\ref{eq:pw_bcs}). (b) Displacement-velocity phase plane for the beam tip, showing regions based on boundary conditions: regions 2 and 6 correspond to inertial forces from tip mass acceleration; regions 3 and 7 include inertial and restoring spring forces; regions 4 and 8 include inertial, spring, and kick forces. If an orbit enters region 8 (or 4) from region 7 (or 3) instead of region 6 (or 2), the kick force is not activated, extending region 7 (or 3). Models A, B, and C describe the dynamics in regions 2/6, 4/8, and the remaining regions, respectively (see Eq.~\ref{eq:weak_eom_cases}).}
    \label{fig:experimental_setup}
\end{figure}

The main structure of the oscillator consists of a Bernoulli-Euler beam with an attached end mass, as depicted in \cref{fig:es_b}. 
A position-triggered piecewise constant force is applied at the tip during oscillations when the horizontal position of the tip mass is within \(\pm d/2\) of the origin (the vertical equilibrium position), provided the tip velocity surpasses a specified critical threshold value, $v_{cr}$. 
Here, \(d/2\) represents a zone of influence within which, an additional restoring force acts on the tip mass: as the tip moves beyond \(\pm d/2\), the restoring force decays rapidly to zero. 
We model this transition with boundary conditions that are piecewise and, therefore, nonsmooth.
More details follow.

The dimensionless governing equation for the Bernoulli-Euler beam  is given by
\begin{equation}
    w^{\prime\prime\prime\prime}(x,t)+\ddot{w}(x,t)+c_{m}\dot{w}^{\prime\prime\prime\prime}(x,t)+c_{v}\dot{w}(x,t)=0, 
    \label{eq:goveq}
\end{equation}
where: $w(x,t)$ ($x \in [0,1]\triangleq \Omega $) denotes the horizontal deflection of the beam as measured from the vertical neutral axis; the primes and overdots indicate spatial and temporal derivatives, respectively;  and $c_m$ and $c_v$ denote the  coefficients of material and viscous damping, respectively (see Appendix~\ref{sec:AppC_ND_derv} for rescaling). 
The boundary conditions for \cref{eq:goveq} are modeled  using the position-velocity phase plane of the beam's tip at $x=1$, as illustrated in \cref{fig:phse}.

The model possesses eight distinct regions with different boundary conditions at $x=1$  because of the different forcing conditions  mentioned above, each of which we now describe.
In regions 2 and 6, there is no restoring force acting on the tip mass $m$. 
As a result, the beam at $x=1$ is only subjected to the inertial force arising from the tip mass. 
In regions 3 and 7, there is no kick force because the tip velocity fails to exceed the critical velocity needed to trigger it (that is, $|\dot{w}(1,t)| < v_{cr}$); however, the restoring force acting on the tip mass is active. 
We model this force by a linear spring from the tip to ground (\cref{fig:experimental_setup}\subref{fig:es_b}). 
As a result, in regions 3 and 7, the boundary condition at $x=1$ includes both the inertial and generalized restoring forces. 
The same is true for regions 1 and 5. 
In regions 4 and 8, the tip velocity magnitude $| \dot{w}(1,t) | > v_{cr}$, so the system is subjected to a force boundary condition, such as might result from an  electromagnet ``kick'' in the motivating physical example. 
We here model this kick as a constant force at the beam tip. We further impose the condition that if the beam's tip trajectory enters region 8 (or 4) from region 7 (or 3), instead of via region 6 (or 2), the kick force is not activated, so that  region 8 (or 4) has the same boundary condition as in region 7 (or 3).

The above boundary conditions can then be summarized as follows (refer to \cref{fig:phse}):
\begin{subequations}
\begin{align}
	& w(0,t) =0\\
	& w'(0,t) =0\\
	& {w}''(1,{t})+{c}_{m}\dot{{w}}''(1,{t}) =0\\
    & {w}'''(1,{t})+{c}_{m}\dot{{w}}'''(1,{t}) =
	\begin{cases} 
		{m}\ddot{w}(1,t) &\text{in regions 2, 6} \\
		{k}w(1,t)+{m}\ddot{w}(1,t)+{F}\text{sgn}(\dot{w}(1,t))&\text{in regions 4, 8}\\
		{k}w(1,t)+{m}\ddot{w}(1,t) &\text{in regions 1, 3, 5, 7}
	\end{cases}\label{eq:pw_bcs_cases}
\end{align}\label{eq:pw_bcs}
\end{subequations}

\vspace{-1\baselineskip}
in which: $m$ denotes the end mass; $k$ denotes the  stiffness of the spring modeling piecewise restoring force at the beam tip, and $F$ is the dimensionless kick force strength generated (see Appendix~\ref{sec:AppC_ND_derv}). Note that in regions 4 and 8 (Eq.~\ref{eq:pw_bcs_cases}) the kick force $F$ acts in same direction as the tip velocity.

These boundary conditions result in three linear models that we label as A, B, and C, that, together, describe the nonlinear dynamics of the oscillator in its hybrid phase space. 
Model A describes the behavior of the system in regions 2 and 6, when the tip is far from the equilibrium position, so there is no tip force (refer to \cref{fig:phse} and the cases of  Eq.~\ref{eq:pw_bcs_cases}). 
The other two models incorporate the effect of the localized, piecewise restoring force: model B describes the behavior in regions 4 and 8, in which both the kick force and restoring force are active; model C describes the dynamics of the beam in regions 1, 3, 5, 7, in which only the restoring force is active.  

\subsection{Weak formulation of the governing equations}
\label{sec:weakForm}
\Cref{eq:goveq}, along with the boundary conditions of \cref{eq:pw_bcs}, represents the strong from of the piecewise initial boundary value problem (IBVP) in the displacement $w$. We aim to determine the solution of the IBVP using the Ritz method \cite{surana_finite_2016} by first deriving the system's weak form \cite{hughes_finite_2000}. 
This solution will thereafter be taken as representing the ``exact'' dynamics of the system.

Let $q(x)$ be selected from a space of trial functions satisfying the essential boundary conditions and having square-integrable spatial derivatives up to second order. We then express the weak form of the IBVP as: 
\begin{equation}
\left(\,\left[\ddot{w}+w''''+c_{v}\dot{w}+c_{m}\dot{w}''''\right],q\,\right) = 0
\label{eq:ip_gov}
\end{equation}
where $(\cdot,\cdot)$ denotes the standard inner product in $L^{2}$. Integrating by parts and employing the boundary conditions we obtain
\begin{multline}
	\int_{0}^{1}\ddot{w}(x,t)q(x)\,dx+c_{v}\int_{0}^{1}\dot{w}(x,t)q(x)\,dx\, + \\
	c_{m}\int_{0}^{1}\dot{w}''(x,t)q''(x)\,dx
	+\int_{0}^{1}w''(x,t)q''(x)\,dx+ \mathcal{I} \,=\, 0
	\label{eq:weak_eom}
\end{multline}

Where $\mathcal{I}$ is defined as (refer to \cref{fig:phse} and \cref{eq:pw_bcs_cases}):
\begin{subnumcases}{\label{eq:weak_eom_cases}\mathcal{I} =} 
		{m}\ddot{w}(1,t)q(1) & \text{for Model A} \label{eq:weak_eom_A}\\
		\left(\,{k}w(1,t)+{m}\ddot{w}(1,t)+{F}\text{sgn}(\dot{w}(1,t))\, \right)q(1) & \text{for Model B} \label{eq:weak_eom_B}\\
		\left(\, {k}w(1,t)+{m}\ddot{w}(1,t)\, \right )q(1) & \text{for Model C}\,. \label{eq:weak_eom_C}
\end{subnumcases}

To derive an approximate numerical solution to these weak IBVPs, one chooses a suitable basis for the space of trial functions, and writes $w(x,t)$ as an expansion in terms of $N$ of the basis functions. Substituting this expansion into \cref{eq:weak_eom} transforms our weak IBVP into a piecewise initial value problem (IVP) with a system of $N$ second-order ODEs valid in each region of the phase space. The solutions to this piecewise system provide the coordinates of each basis function needed to determine and approximation to $w(x,t)$. In principle, this can be done using  either a localized basis, as with finite element analysis, or a global basis, as with modal analysis. Owing to the relatively simple structure of our system, we elected to use the latter, as described in the next section.

\subsection{Piecewise solution to the weak boundary value problem\label{sec:pw}}

To obtain the linear normal modes for each model, we first estimate the associated natural frequencies. 
In all cases, the characteristic equation can be written in the general form: 
\begin{equation}
	\label{eq:characteristic_eq}
	(\mathscr{A}-\mathscr{B})\exp(-2\text{\ensuremath{\beta}})+(\mathscr{A}+\mathscr{B})+2\text{\ensuremath{\beta}}^{3}\exp(-\text{\ensuremath{\beta}})=0,
\end{equation}
where
\begin{subequations}
	\begin{equation}
		\mathscr{A}  =  \beta^{3}\cos(\beta)+{k}\sin(\beta)+\beta^{4}(-{m})\sin(\beta)\label{eq:A},\\
	\end{equation}
	\begin{equation}
		\mathscr{B}  =  \beta^{4}{m}\cos(\beta)-{k}\cos(\beta)\label{eq:B},
	\end{equation}
	\label{eq:characteristic_eq_AB}
\end{subequations}

\vspace{-1.5\baselineskip}
and $\beta=\sqrt{\omega}$, where $\omega$ is a natural frequency of the beam. 

The characteristic equation \cref{eq:characteristic_eq} is the same for models B and C because the constant force ${F}$ at the boundary in model B can be eliminated by first calculating the solution of its PDE about the static deflection curve resulting from ${F}$ (see Appendix \ref{sec:asimub}). The natural frequencies in these cases are denoted as by $\omega_i$ ($i=1,2,...,N$).
For model A, \cref{eq:characteristic_eq,eq:characteristic_eq_AB} are modified by setting ${k}=0$ and the natural frequencies are denoted by $\Omega_i$.

\Cref{eq:characteristic_eq} was used to numerically solve for the natural frequencies, which were then used to obtain the normal modes \cite{meirovitch_fundamentals_2001,Rao2006}. 
Denoting the $i^{th}$ mode for model A as $\gamma_i$, and for models B and C as $\xi_i$, $N$-mode approximations for $w(x,t)$ have the form:
\begin{subnumcases}
{\label{eq:mod_w_ABC}w(x,t)=}
	\sum_{i=1}^{N} r_i(t)\gamma_i(x) & for model A \label{eq:mod_w_A} \\
	\sum_{i=1}^{N} a_i(t)\xi_i(x) & for model B \label{eq:mod_w_B} \\
	\sum_{i=1}^{N} \alpha_i(t)\xi_i(x) & for model C. \label{eq:mod_w_C}
\end{subnumcases}
Starting with the eigenvalue problem in each case, standard manipulations (the general form of which can be found in Appendix~\ref{sec:AppC1}) lead to the following orthonormality conditions:
\begin{subequations}
\begin{gather}
		\int_{0}^{1}\gamma_{i}\gamma_{j}dx+{m}\gamma_{i}(1)\gamma_{j}(1)=\delta_{ij},\label{eq:normalize_A_KE}\\
		\int_{0}^{1}\gamma''_{j}\gamma''_{i}dx=\Omega^2_{i}\delta_{ij},\label{eq:normalize_A_PE}\\
		\int_{0}^{1}\xi_{i}\xi_{j}dx+{m}\xi_{i}(1)\xi_{j}(1)=\delta_{ij},\label{eq:normalize_BC_KE}\\
		\int_{0}^{1}\xi''_{i}\xi''_{j}dx + k \xi_{i}(1)\xi_{j}(1)=\omega^2_{i}\delta_{ij}.\label{eq:normalize_BC_PE}
\end{gather}\label{eq:normalize_ms}
\end{subequations}

\vspace{-1\baselineskip}
Using  the computed normal modes in the expansions of Eqs.~(\ref{eq:mod_w_ABC}), substituting into  \cref{eq:weak_eom}, and integrating yields coupled ordinary differential equations (ODEs) for each model. For example, with model C, Eq.~\eqref{eq:mod_w_C} yields
\begin{multline}
	\sum_{i=1}^{N}\left\{ \ddot{\alpha_{i}}\int_{0}^{1}\xi_{i}\xi_{j}dx+\alpha_{i}\int_{0}^{1}\xi''_{i}\xi''_{j}dx+c_{v}\dot{\alpha_{i}}\int_{0}^{1}\xi_{i}\xi_{j}dx +c_{m}\dot{\alpha_{i}}\int_{0}^{1}\xi''_{i}\xi''_{j}dx \right . \\
	\left. +\big ({k}\alpha_{i}\xi_{i}(1)+{m}\ddot{\alpha_{i}}\xi_{i}(1)\big )\,\xi_{j}(1) \vphantom{\Big |} \right\}  = 0
\end{multline}
Then, Eqs.~\eqref{eq:normalize_BC_KE} and \eqref{eq:normalize_BC_PE} yield $N$ ODEs as:
\begin{align}
		& \sum_{i=1}^{N}\left\{ \delta_{ij}(\ddot{\alpha_{i}}+\omega_{i}^{2}\alpha_{i})+c_{v}\dot{\alpha_{i}}\int_{0}^{1}\xi_{i}\xi_{j}dx+c_{m}\dot{\alpha_{i}}\int\xi''_{i}\xi''_{j}dx\right\}  = 0 \notag\\
\implies &\; \ddot{\alpha}_{j}+
\left(c_{v}+c_{m}\omega_{j}^{2}\right)\dot{\alpha}_{j}+\sum_{i=1}^{N} D_{ij} \dot{\alpha}_{i}+\omega_{j}^{2}\alpha_{j}  = 0\quad(j = 1, 2,\ldots,N),\label{eq:coupled_odes}
\end{align}
in which the matrix with elements $D_{ij}=-\xi_{i}(1)\xi_{j}(1)({m}c_{v}+{k}c_{m})$ represents non-proportional damping arising from the $x=1$ boundary condition. Hence, the system is not diagonalizable: that is, all of its normal modes are coupled through the damping. The ODEs for Model B have the same form as \cref{eq:coupled_odes} for displacements relative to the static deflection curve for constant $F$, as described above. Similarly, using Eqs.~\eqref{eq:normalize_A_KE} and \eqref{eq:normalize_A_PE} one obtains the ODEs for model A, which have the same form as \cref{eq:coupled_odes}, but with ${k}=0$, normal modes $\xi_i$ replaced by $\gamma_i$, and natural frequencies $\omega_i$ replaced by $\Omega_i$.

At the beginning of a simulation, a specific model is chosen depending on where the initial position and velocity of the tip of the beam is located. This is used to obtain the dynamics of the beam until its tip hits a transition boundary, either at $w(1,t)=0$ or $|w(1,t)|=d/2$  (Figs.~\ref{fig:experimental_setup}\subref{fig:es_b} and \ref{fig:phse}). The state at the transition is transferred to the new region's model and subsequent dynamics is computed.

For example, consider that the initial configuration of the beam is such that the tip of the beam is in region 2 (\cref{fig:phse}). In this case, model A is used to describe the dynamics of the beam. As the system state evolves in time, the trajectory eventually enters region 3 or 4 in the phase plane, depending on the velocity of the tip. 
The time $t_R$ at which the tip of the beam reaches the transition boundary is estimated from $|w(1,t_R)| =d/2$ using model A and the bisection root finding algorithm. 
Across the boundary, the full state of model A must be passed to the next model.  
Hence, assuming, for example, that the beam tip enters region 3, at the transition from model A to model B at $t=t_{R}$, the following must hold:
\begin{equation}
\begin{aligned}
w(x,t_{R})   &= \sum_{i=1}^{N} r_i(t_{R})\gamma_i(x)= \sum_{j=1}^{N} a_j(0)\xi_j(x)\\
\dot{w}(x,t_{R}) &= \sum_{i=1}^{N} \dot{r}_i(t_{R})\gamma_i(x) = \sum_{j=1}^{N} \dot{a}_j(0)\xi_j(x).
\end{aligned}\label{eq:field_transition}
\end{equation}
Note that we can take $t=0$ in the post-transition model because the system is autonomous. This implies that
\begin{equation}
\begin{aligned}
a_i(0) &=  \sum_{j=1}^{N}\text{G}_{ij}r_j(t_{R})\\
\dot{a}_i(0) &=  \sum_{j=1}^{N}\text{G}_{ij}\dot{r}_j(t_{R})
\end{aligned}\label{eq:amplitude_transition}
\end{equation}
where $\text{G}_{ij}=\int_{0}^{1}\gamma_i(x)\xi_j(x)\,dx+{m}\gamma_i(1)\xi_j(1).$ 
These initial conditions for the new modal coordinates are then used to calculate the subsequent dynamics using model B. This same approach is adopted for dealing with other similar transitions for the duration of the simulation.

\section{Dynamics of the kicked oscillator}
\label{sec:Sr}
\begin{figure}[t!]
    \centering
    \includegraphics[width=1\textwidth]{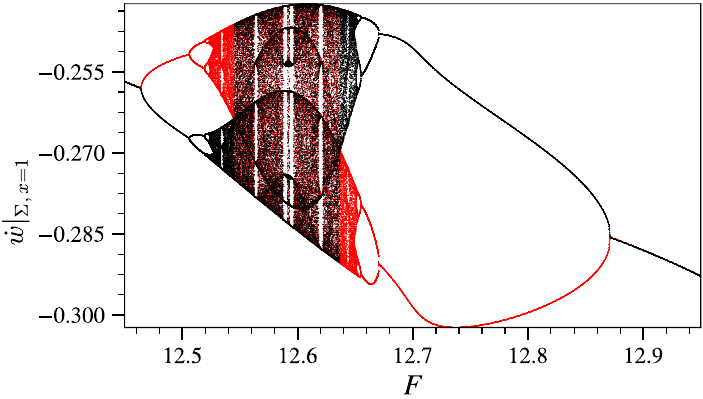}
    \caption{Bifurcation diagram of the kicked oscillator obtained from simulating the piecewise-linear system of \cref{eq:weak_eom,eq:weak_eom_cases} for kick strength $F\in [12.45,12.95]$. ``Exact'' solutions were computed using Ritz approximations with $N=25$ linear normal modes (\cref{sec:pw}). The diagram was constructed on the Poincar\'{e} section $\Sigma$ (Eq.~\ref{eq:section}) using branch-following continuation. The other parameter values used in the simulations were: $c_v=4.5$,  $c_m=3\times 10^{-4}$, $m=1$, $k=1000$,  $d=0.2$, $v_{cr}=0.05$. Two solution branches are shown in the figure: the red branch was obtained by starting at $F=12.95$ and \textit{decreasing} the kick force, whereas the black branch was obtained by starting at $F=12.45$ and \textit{increasing} its value.}
\label{fig:bfd}
\end{figure}

After initial experimentation, it was found that the dynamics of the the hybrid system could be simulated to high fidelity using the expansions of Eqs.~\eqref{eq:mod_w_ABC} with $N=25$ normal modes for all three models (Eqs.~\ref{eq:weak_eom} and \ref{eq:weak_eom_cases}). Thus, a total of 75 different basis functions were used across the system's hybrid phase space. These simulations provided the data for the ``exact'' system. 

MATLAB's native solver, $\texttt{ode45}$, was used to integrate the resulting ordinary differential equations. 
After initial experimentation, we selected ${m}=1$, $d=0.2$, and $v_{cr} = 0.05$. We also set ${k}=1000$ (the ratio of the stiffness to ground at the tip vs. the internal stiffness of the beam), $c_v=4.5$ (the coefficient of viscous damping), and $c_m=3\times 10^{-4}$ (the coefficient of material damping) for simulations. Keeping these parameters fixed for all simulations, we varied the kick strength ${F}$ and recorded various responses of the system. 
We collected discrete samples in space and time of the simulated response, $w(x,t)$, at $N_{\Omega}=100$ uniformly spaced points along the beam, with a sampling frequency $f_s = 1000$ samples/unit time.

Different attracting steady-state solutions of the system are summarized by the bifurcation diagram of \cref{fig:bfd}, in which the kick strength $F$ is the bifurcation parameter. The bifurcation diagram is constructed using a fixed Poincar\'{e} section \cite{guckenheimer1983nonlinear}, defined by
\begin{equation}\label{eq:section}
\Sigma = \big \{\, (w(x,t),\dot{w}(x,t)) \,\,|\,\, w(1,t)=d/2 \,;\,\;\dot{w}(1,t)< 0\, \big \}.
\end{equation}
For each value of $F$, the tip velocity, $\dot{w}(1,t)$ of the steady state at each intersection with $\Sigma$ is plotted. Period-$n$ solutions have $n$ distinct points for the associated value of $F$, whereas chaotic solutions have a theoretically infinite number of points: for \cref{fig:bfd}, we plotted a maximum of 16 distinct points for each $F$. Only dynamic steady states are shown; not shown is the trivial solution (the static equilibrium with $\dot{w}|_{\Sigma, x=1} = 0$), which always exists and has a small basin of attraction determined by $v_{cr}$ (Fig.~\ref{fig:phse}).

Starting with $F = 12.95$, the entire bifurcation diagram was constructed using simple branch-following continuation \cite{seydel_practical_1994} with 
 $\Delta F = 5\times 10^{-5}$, resulting in a total of $10^4$ values of $F$ on the bifurcation diagram. 
The solution branch thus obtained is shown in red in \cref{fig:bfd}. 
The other solution branch shown using black was obtained by starting the same algorithm from $F = 12.45$ and increasing $F$ until $F = 12.95$ with the same $\Delta F$. 

\begin{figure}[tbp!]
    \centering
    \begin{subfigure}[b]{0.48\linewidth}
    \includegraphics[width=\linewidth]{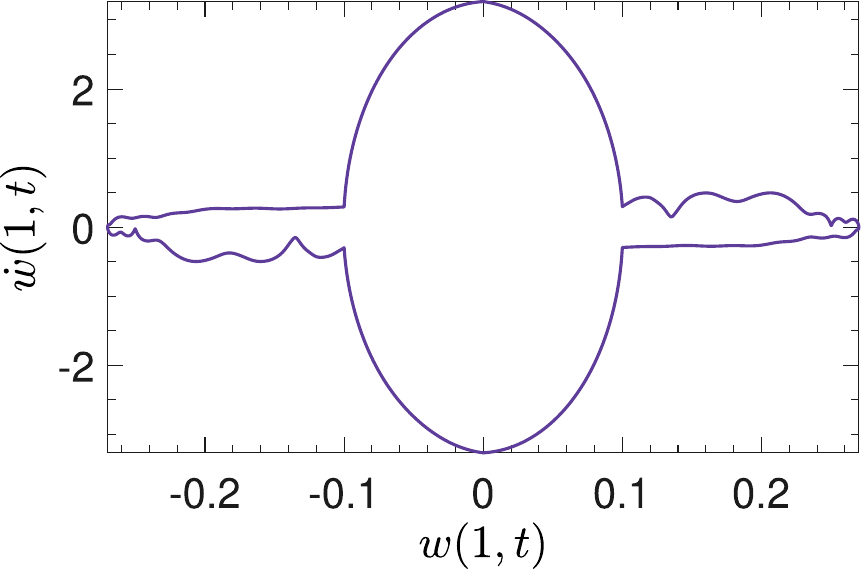}
    \caption{}
    \label{fig:pp_a}
    \end{subfigure}
	\hfill
    \begin{subfigure}[b]{0.48\linewidth}
    \includegraphics[width=\linewidth]{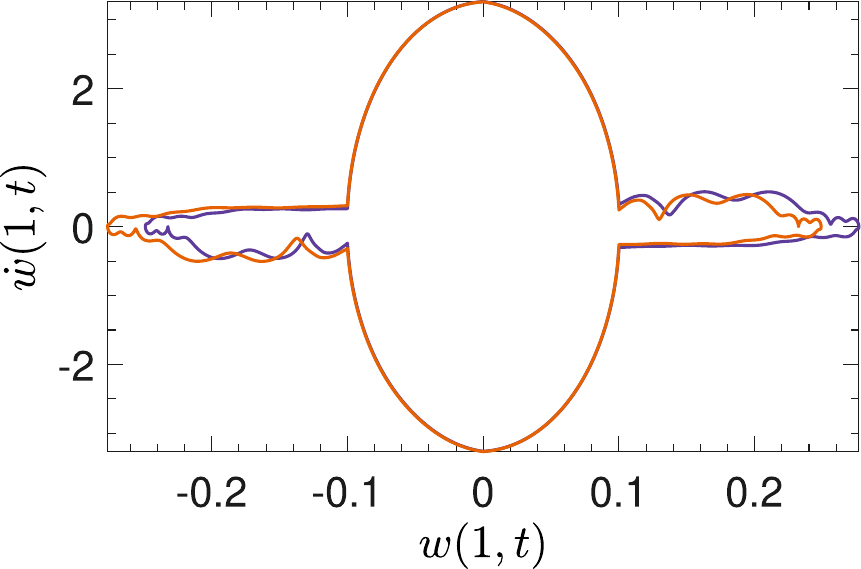}
    \caption{}
    \label{fig:pp_b}
    \end{subfigure}\\[-4pt]
    \begin{subfigure}[b]{0.48\linewidth}
    \includegraphics[width=\linewidth]{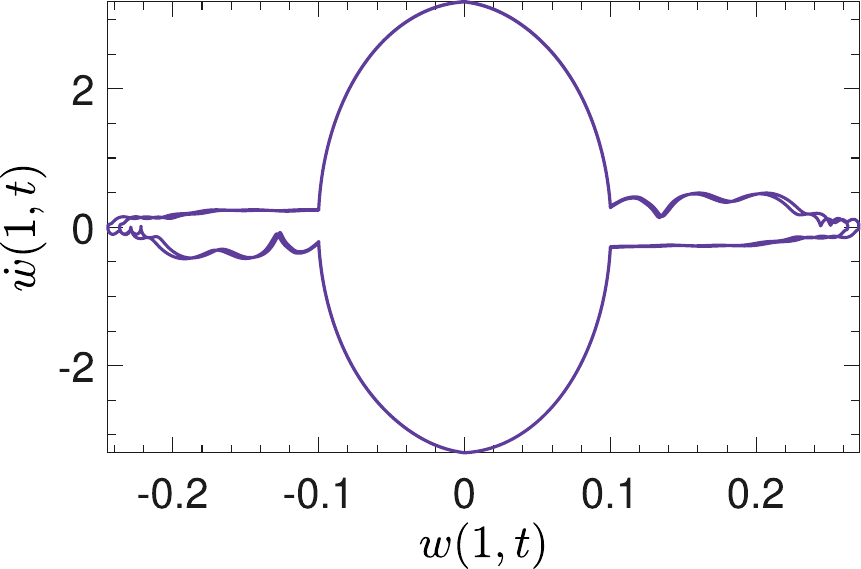}
    \caption{}
    \label{fig:pp_c}
    \end{subfigure}
    \hfill
    \begin{subfigure}[b]{0.48\linewidth}
    \includegraphics[width=\linewidth]{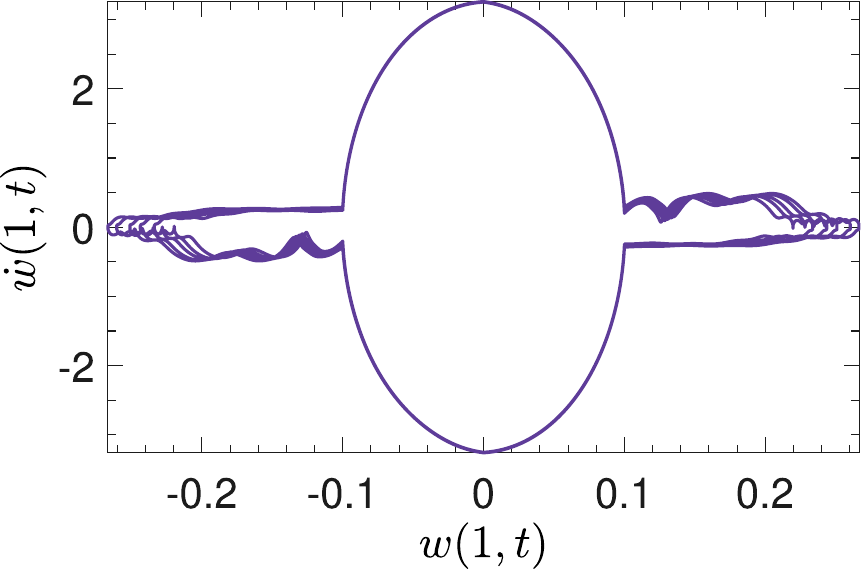}
    \caption{}
    \label{fig:pp_d}
    \end{subfigure}\\[-4pt]
    \begin{subfigure}[b]{0.48\linewidth}
    \includegraphics[width=\linewidth]{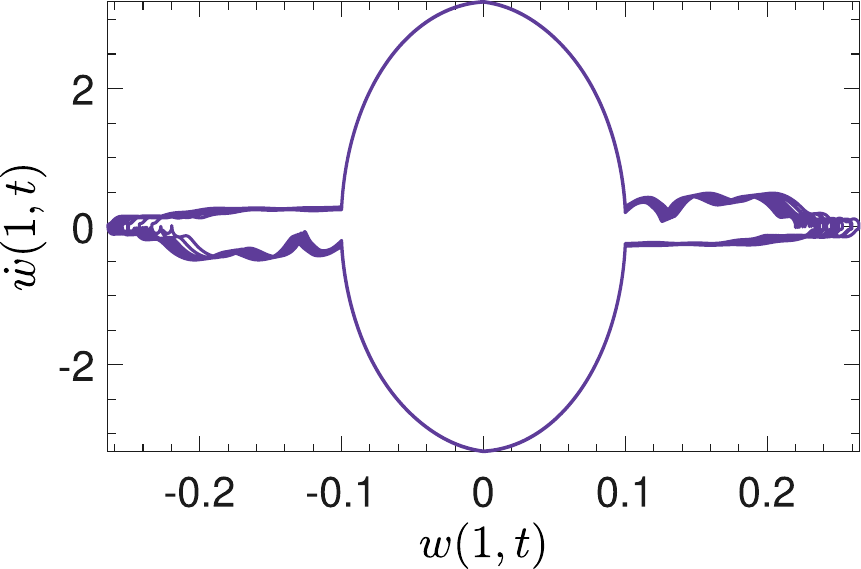}
    \caption{}
    \label{fig:pp_e}
    \end{subfigure} 
    \vspace{-8pt}
    \caption{Different steady state solutions plotted on the phase plane of the beam tip for different kick strengths $F$ (see bifurcation diagram \cref{fig:bfd}): (a) period-1 limit cycle for $F\approx 12.95$; (b) two period-1 limit cycles from different solution branches (indicated by red and black in \cref{fig:bfd}) for $F\approx 12.75$; (c) period-2 orbit for $F\approx 12.66$; (d) period-5 orbit for $F\approx 12.621$; (e) chaotic orbit for $F\approx 12.605$. All orbits, except for those in (b), are symmetric when reflected across both the $w(1,t)$ and $\dot{w}(1,t)$ axes, consistent with the invariance system (Eqs.~\ref{eq:goveq} and \ref{eq:pw_bcs}) under $w\rightarrow -w$ and $\dot{w}\rightarrow -\dot{w}$.  The two asymmetric limit cycles in (b), taken together, recover full symmetry of the system. Except for $F$, the remaining parameters used to construct these plots are the same as used for \cref{fig:bfd}.}
\end{figure}

In \cref{fig:bfd}, we observe that for $F \gtrsim 12.88 $, all trajectories converge to the same period-1 limit cycle: both solution branches are merged in this regime. 
In \cref{fig:pp_a}, the limit cycle corresponding to $F\approx 12.95$ is plotted on the phase plane at the beam's tip. 
We observe that the limit cycle is symmetric under reflections across both the $w(1,t)$ and $\dot{w}(1,t)$ axis. 
This is not surprising, since the governing equations possesses this symmetry, which is evident from the invariance of the PDE (Eq.~\ref{eq:goveq}) and boundary conditions (Eq.~\ref{eq:pw_bcs}) under the transformation $w\rightarrow -w$ and $\dot{w}\rightarrow -\dot{w}$. 
The shape of the orbit is oval-shaped when $|w(1,t)| \leq d/2 = 0.1$, in the kicking region and the trajectory has little or no ``wiggles'' related to higher-frequency components of the oscillations, consistent with only a few modes being excited in this region. 
Immediately outside this region, however, where the linear restoring force and kick force sharply drop to zero, more ``wiggles'' are observed in the ``wing'' shaped segments of the orbit, as the tip crosses from region 4 (or 8) to 5 (or 1). This suggests that the nonsmooth transition has excited a larger number of modes.  Nondifferentiable ``kinks'' in the orbits occur in two ways: the more obvious case occurs when all forces at the tip suddenly drop to zero whenever $|w(1,t) |$ exceeds $d/2$; the more subtle case occurs when the kick force suddenly turns off as $w(1,t)$ crosses zero in either direction.

As the value of $F$ is decreased, the amplitude of the period-1 limit cycle decreases until $F\approx 12.88$, at which we observe two period-1 solution branches emerge and continue until $F\approx 12.68$. This does \textit{not} indicate doubling but, rather, a symmetry-breaking bifurcation in which two, distinct asymmetric period-1 orbits are created. 
In \cref{fig:pp_b}, we plot both of these orbits from both branches, corresponding to $F\approx 12.75$: unlike in \cref{fig:pp_a}, these individual solutions do not possesses  reflectional symmetry across the plane's axes; however, taken together, they recover the full system symmetry. 

Below $F\approx 12.68$, we observe period-doubling bifurcation sequences in both branches. 
In \cref{fig:pp_c}, we show a period-2 orbit appearing after the first period doubling in the lower (red) branch at $F\approx 12.66$. 
As $F$ is further decreased, the period doubling both branches leads to chaotic regimes. Eventually, for $12.545 \lesssim F \lesssim 12.635$, both solution branches merge and lead to a single chaotic regime with full symmetry, with brief windows containing multi-period solutions. 
In \cref{fig:pp_d,fig:pp_e}, we show a period-5 and a chaotic orbit on the phase plane of the beam tip, corresponding to $F\approx 12.621$ and $F\approx 12.605$, respectively. 
For $F\lesssim12.545$, we again observe two different solution branches undergoing period halving bifurcations separately and merging into a single branch for $F\lesssim 12.52$. This branch abruptly ends at $F \approx 12.41$ as the system no longer converges to a dynamic steady state. Instead, the vertical static equilibrium position becomes globally attracting. This occurs because it is no longer possible for $| \dot{w}(1,t) |$ to exceed $v_{cr}$, so the kick is never activated. 

\section{Model reduction procedure}
\label{sec:mr}
   
Proper orthogonal decomposition (POD) starts with the continuous displacement field $w(x,t)$ and  finds the basis functions $\psi(x)$  such that the time-averaged projection error of $w$ onto $\psi$ is minimized \cite{holmes_turbulence_1996,Bhattacharyya_2020} :
\begin{equation}
	\underset{ \psi_i \in L^2(\Omega)}{\operatorname{min}} \Big \langle \,\big\| w(x,t)-\sum_{i=1}^{P}(w(x,t),\psi_i)\mspace{1mu}\psi_i \, \big\|^{2} \Big \rangle,
	\label{eq:conti}
\end{equation}
in which angle brackets $\langle\cdot\rangle$ denote the time averaging operation, $\Omega=[0,1]$, and the objective function is subjected to the normality constraint $\|\psi\|=1$. 

It can be shown \cite{holmes_turbulence_1996} that the above minimization problem is equivalent to the following infinite-dimensional eigenvalue problem:
\begin{equation}
	\int_{\Omega}r(x,y)\psi(y)\,dy=\lambda \psi(x),
	\label{eq:EVP_cont}
\end{equation}
where the integral kernel $r(x,y)$ is the spatial correlation function for $x,y \in \Omega$: 
\begin{equation}
	r(x,y)=\left\langle\, w(x,t)\mspace{2mu}w(y,t)\,\right\rangle.
	\label{eq:acf}
\end{equation}
Solving the eigenvalue problem \cref{eq:EVP_cont} leads to an infinite set of eigenfunctions $\{\psi_i\}_{i=1}^\infty$, in this case referred to as proper orthogonal modes (POMs), each with a corresponding eigenvalue $\lambda_i$ that denotes the variance of $w(x,t)$ along $\psi_i(x)$. 
The POMs are sorted based on the magnitudes of the associated $\lambda_i$ and the first $P<N$ of these modes are chosen to form a $P$-dimensional subspace, onto which the governing equation of the system is projected. The model reduction problem then becomes essentially that of identifying a ``good'' value of $P$.

Conventionally, $P$ is selected such that the reduced subspace spanned by \(\{\psi_i(x)\}_{i=1}^{P}\) captures a significant fraction of the \textit{variance} of \(w\). 
In practice, this is achieved by ensuring that the uncaptured variance
\begin{equation}
    e_{_P} = 1 - \frac{\sum_{i=1}^P \lambda_i}{\sum_{i=1}^N \lambda_i}
    \label{eq:e_k}
\end{equation}
remains below a prescribed tolerance \(\varepsilon\); typically \(\varepsilon < 10^{-3}\) \cite{sirovich_chaotic_1989}.

For the current problem, the displacement and velocity data obtained from simulations of the hybrid system described by models A, B, and C (Eqs.~\ref{eq:weak_eom} and \ref{eq:weak_eom_cases}), were discretely sampled and collected in matrices $\mat{W}, \dot{\mat{W}} \in \mathbb{R}^{N_{\Omega}\times N_{t}}$ such that
\begin{subequations}\label{eq:W_data}
\begin{align}
	W_{ij} &= w(x_i,t_j) \label{eq:W_ij}\\
	{\dot{W}}_{ij} &= \dot{w}(x_i,t_j)\label{eq:Wdot_ij}\,,
\end{align}
\end{subequations}
where $N_{\Omega}$ and $N_{t}$ denote the number of discrete measurements in the spatial and temporal domains, respectively.  Thus, the correlation kernel $r(x,y)$ of \cref{eq:acf} becomes the matrix 
\begin{equation}
	\mat{R}= \frac{1}{N_{t}}\mat{W}\mat{W}^\intercal \in \mathbb{R}^{N_{\Omega}\times N_{\Omega}},
	\label{eq:disc_cov}
\end{equation}
in which trapezoidal integration is used to approximate the time-average \cite{Brunton_2019}. Eigenvectors $\boldsymbol{\uppsi}\in \mathbb{R}^{N_{\Omega} \times 1}$ from the matrix eigenvalue problem
\begin{equation}
	\mat{R}\boldsymbol{\uppsi}=\lambda\boldsymbol{\uppsi}
	\label{eq:disc_EVP}
\end{equation}
are interpolated using Chebychev polynomials to yield functions $\widehat{\psi}(x)$ continuous in $x$ that approximate the eigenfunctions $\psi(x)$ (that is, the POMs) in \cref{eq:EVP_cont}. The displacement field is then approximated 
by a Ritz expansion using the POMs:
\begin{equation}
	w(x,t) \approx \widehat{w}(x,t)=\sum_{j=1}^{P}b_{j}(t)\widehat{\psi}_{j}(x).
	\label{eq:romd}
\end{equation}

The data used for the above calculations was collected as the system traversed all 8 regions in the phase space (Fig. \ref{fig:phse}). We thus obtain one set of POMs for the entire hybrid phase space: these are used to construct a $P$-dimensional piecewise linear ROM. Having a global set of POMs obviates the need to separately ensure continuity for the displacement and velocity fields across transition boundaries using different basis functions, as was necessary during simulations via Eqs.~\eqref{eq:field_transition} and \eqref{eq:amplitude_transition}. 

Substituting the expansion \cref{eq:romd} in terms of the POMs into the weak form of the system's governing equations, Eqs.~\eqref{eq:weak_eom} and \eqref{eq:weak_eom_cases}, and performing a Galerkin projection on the residual, we obtain a $P$-dimensional system of second order ODEs for each of models A, B, and C (Sec.~2). For example, consider model C, for which the boundary conditions Eq.~\eqref{eq:weak_eom_C} contain both inertial and restoring forces. These  yield the following set of $P$ coupled ordinary differential equations (ODEs) in terms of the proper orthogonal coordinates $b_{j}(t)$:
\begin{multline*}
	\sum_{i=1}^{P} \left \{\ddot{b}_{i}(t)\int_{0}^{1}\widehat{\psi}_{i}(x)\widehat{\psi}_{j}(x)dx+b_{i}(t)\int\widehat{\psi}''_{i}(x)\widehat{\psi}''_{j}(x)dx+c_{v}\dot{b}_{i}(t)\int_{0}^{1}\widehat{\psi}_{i}(x)\widehat{\psi}_{j}(x)dx \right .\\
	\left . +c_{m}\dot{b}_{i}(t)\int\widehat{\psi}''_{i}(x)\widehat{\psi}''_{j}(x)dx+\left(kb_{i}(t)+{m}\ddot{b}_{i}(t)\right)\widehat{\psi}_{i}(1)\widehat{\psi}_{j}(1) \right \}  =0
\end{multline*}
\begin{equation}
	\implies\quad \sum_{i=1}^{P}\left \{ \big [M_{ij}+m{E}_{ij} \big ]\,\ddot{b}_{i}(t)+\big [ M_{ij}c_{v}+c_{m}K_{ij}\big ]\, \dot{b}_{i}(t)+\big [K_{ij}+k{E}_{ij}\big ] \,b_i(t) \right \}  =0,
	\label{eq:ROM}
\end{equation}
where 
\begin{equation}
		M_{ij}=\int_{0}^{1}\widehat{\psi}_{i}(x)\widehat{\psi}_{j}(x)dx
\quad\text{and}\quad
K_{ij}=\int_{0}^{1}\widehat{\psi}''_{i}(x)\widehat{\psi}''_{j}(x)dx
\end{equation}
are the main elements of the mass and stiffness matrices, respectively, and
\begin{equation}
		{E}_{ij}=\widehat{\psi}_{i}(1)\widehat{\psi}_{j}(1)
\end{equation}	
are elements of a coupling matrix arising from the boundary condition at $x=1$: examination of \cref{eq:ROM} indicates that this boundary coupling also contributes the the ROM's overall mass and stiffness properties. Note that, like the original system,  the ROM does not have proportional damping, so it also fails to be diagonalizable. Similar calculations yield systems of ODEs for models A and B.

\begin{figure}[t]
\begin{algorithm}[H]
    \caption{Estimating ROM dimension with POD and energy closure criterion.}
    \label{alg:dim_sel}
    \begin{algorithmic}[1]
        \vspace{6pt}
        
        \STATE \textbf{Input:} $\mat{W}\in\mathbb{R}^{N_{\Omega}\times N_t}$, $\dot{\mat{W}}\in\mathbb{R}^{N_{\Omega}\times N_t}$\vspace{6pt}
        
        \STATE \textbf{Output}: $P_{\,\mathrm{ROM}}$\vspace{6pt}
        
        \STATE \textbf{Set:} tolerance = $\epsilon_{\text{tol}}$ ($10^{-4}$ in our case).\vspace{6pt}
        
        \STATE Calculate $P_{\,\mathrm{max}} = \operatorname{rank}(\mat{W})$\vspace{6pt}
        
        \STATE Perform POD: $\{\uppsi_\text{max}, \boldsymbol{\sigma} \} = \mathrm{POD}(\mat{W})$, where $\uppsi_\text{max}\in\mathbb{R}^{N_{\Omega}\times P_{\,\mathrm{max}}}$ and $\boldsymbol{\sigma}\in\mathbb{R}^{P_{\,\mathrm{max}}\times 1}$ \vspace{6pt}
        
        \STATE Using $\mat{W}$ and $\mat{\dot{W}}$ calculate $\widetilde{W}^{P_\text{max}}_f $ and $\widetilde{W}^{P_\text{max}}_d $ (Eq.~\ref{eq:WdWf}). \vspace{6pt}
        
        \STATE \textbf{Initialize:} $P=P_{\,\text{init}}$, $e_d = e_f= 100\,\epsilon_{\text{tol}}$, $\uppsi = {\uppsi_\text{max}}[\,:\,,\,1\!:\!P\,] $  \vspace{6pt}
        
        \WHILE{$e_d$ or $e_f > \epsilon_{\text{tol}}$ AND $P \le \operatorname{rank}(\mat{W})$}\vspace{6pt}
            \STATE Project full state data onto subspace spanned by ${\uppsi}$ (Eq.~\ref{eq:proj_vel})        \vspace{6pt}
            
            \STATE Calculate $\widetilde{W}_d$ and $\widetilde{W}_f$ (Eq.~\ref{eq:WdWf}).\vspace{6pt}
            
            \STATE Calculate ${e}_d$ and ${e}_f$  (Eq.~\ref{eq:edef}).\vspace{6pt}

            \IF{$e_d$ or $e_f > \epsilon_{\text{tol}}$}\vspace{6pt}
                \STATE increment: $P = P + 1$\vspace{6pt}
                \STATE $\uppsi \leftarrow [\,\uppsi\,,\,\uppsi_\text{max}[\,:\,,P\,]\,]$\vspace{6pt}
            \ELSE\vspace{6pt}
                \STATE $P_{\,\text{ROM}} = P$\vspace{6pt}
                \STATE \text{\bf STOP} \vspace{6pt}
            \ENDIF\vspace{6pt}
        \ENDWHILE
    \end{algorithmic}
\end{algorithm}
\vspace{6pt}
\end{figure}

To this point, our procedure has employed a relatively straightforward implementation of standard POD. 
However, instead of the conventional approach described in \cref{eq:e_k} to select the reduced order model (ROM) dimension $P$, we here adopt the energy closure criterion \cite{Bhattacharyya_2020,Bhattacharyya_2022}, summarized below.

$P$ is chosen to ensure that the subspace spanning $\{\widehat{\psi}_i(x)\}_{i=1}^P$ is closed with respect to energy flowing in and out of it, to within a specified numerical tolerance. 
This approach provides a physics-based explanation for why a given dimension is needed in a way that mere examination of the total variance cannot.

In this case, the dissipation energy (or work), $W_d$, and input energy (or work), $W_f$, for the full-order system over one period $T$ of a steady-state periodic response are given by
\begin{subequations}
	\begin{align}
		{W}_f= & \int_{0}^{T}\,f_{\textrm{kick}}{\dot{w}}(1,t)\,dt\;\;\text{and}\label{eq:Wf}\\ 
		{W}_d = & \,c_v\mspace{-2mu}\int_{0}^{1}\int_{0}^{T}{\dot{w}}^{2}(x,t)\,dt\,dx+\,c_m\mspace{-2mu}\int_{0}^{1}\int_{0}^{T}{\dot{w}}''^{2}(x,t)\,dt\,dx,\label{eq:Wd}\		\end{align}
	\label{eq:WdWf}
\end{subequations}

\vspace{-1\baselineskip}
where $f_{\textrm{kick}}$ is the state-dependent kick force, as defined in Sec.~\ref{sec:md} and summarized in \cref{fig:phse}. 
At steady-state, $W_d$ and $W_f$ must be in balance: 
\begin{equation}
	W_d=W_f.
	\label{eq:wd_e_wf}
\end{equation}
To check whether this is the case on a subspace of dimension $P$, we estimate them using the $P$-dimensional projection of the original velocity data and its spatial derivatives in Eqs.~\eqref{eq:WdWf}. 
We write these estimates as  $\widetilde{W}_d$ and $\widetilde{W}_f$, respectively. For an insufficiently large value of $P$, $\widetilde{W}_d$ and $\widetilde{W}_f$ will not be in balance. 

The projection of the discrete velocity field stored in the matrix $\dot{\mat{W}}$ (Eq.~\ref{eq:Wdot_ij}) is given by
\begin{equation}
	\widetilde{\dot{\mat{W}}} = \mat{\uppsi}\mat{\uppsi}^\intercal{\dot{\mat{W}}}
	\label{eq:proj_vel}
\end{equation}
where $\mat{\uppsi} \in \mathbb{R}^{N_{\Omega}\times P}$ is a matrix with columns consisting of eigenvectors found from \cref{eq:disc_EVP}.

To reduce discretization errors in the work estimates $\widetilde{W}_d$ and $\widetilde{W}_f$, which depend on the velocity and strain rate fields (Eq.~\ref{eq:WdWf}), we interpolated each column of $\widetilde{\dot{\mat{W}}}$, representing the spatial component of the velocity data at every time sample, using Chebychev polynomials, as was done to interpolate the POMs. This ensured lower error in the estimates of the second order spatial derivative required for \cref{eq:Wd}.
The time integrations needed for both energy estimates \cref{eq:WdWf} were performed using the trapezoidal rule; the spatial integrations were numerically evaluated using the Chebyshev polynomial representation \cite{driscoll2014}.

Given $\widetilde{W}_d$ and $\widetilde{W}_f$, we define the energy convergence errors \cite{Bhattacharyya_2022},
\begin{equation}\label{eq:edef}
{e}_f = \left| 1 - \frac{\widetilde{W}^{\hphantom{P_\text{max}}}_f}{\widetilde{W}^{P_\text{max}}_f} \right|
\quad\text{and}\quad
{e}_d = \left| 1 - \frac{\widetilde{W}^{\hphantom{P_\text{max}}}_d}{\widetilde{W}^{P_\text{max}}_d} \right|,
\end{equation}
of the input and dissipation energy, respectively: both ${e}_d$ and ${e}_f$ are functions of $P$. In these expressions, the superscript $P_\text{max}$ indicates the energy value estimated at $P=P_\text{max}$, where $P_\text{max}$ is the numerically estimated rank of $\mat{R}$ (Eq.~\ref{eq:disc_cov}). POMs for $P>\text{rank}(\mat{R})$ lie in the null space of $\mat{R}$ and do not capture the system's dynamics \cite{Bhattacharyya_2020}. Hence, $P_\text{max}$ indicates the maximum ROM dimension that a given data set is capable of capturing. Once $\widetilde{W}_f$ and $\widetilde{W}_d$ are captured with sufficient accuracy, physics \textit{requires} that they approximately balance each other. Thus, the value of the subspace dimension $P$ for which ${e}_f$ and ${e}_d$ drop below a given predefined tolerance $\epsilon_{\text{tol}}$, is selected as the ROM dimension (see \cref{alg:dim_sel}).

\section{Results}
\begin{figure}[t!]
\centering
	\begin{subfigure}[b]{0.49\linewidth}
	\includegraphics[height=.8\linewidth]{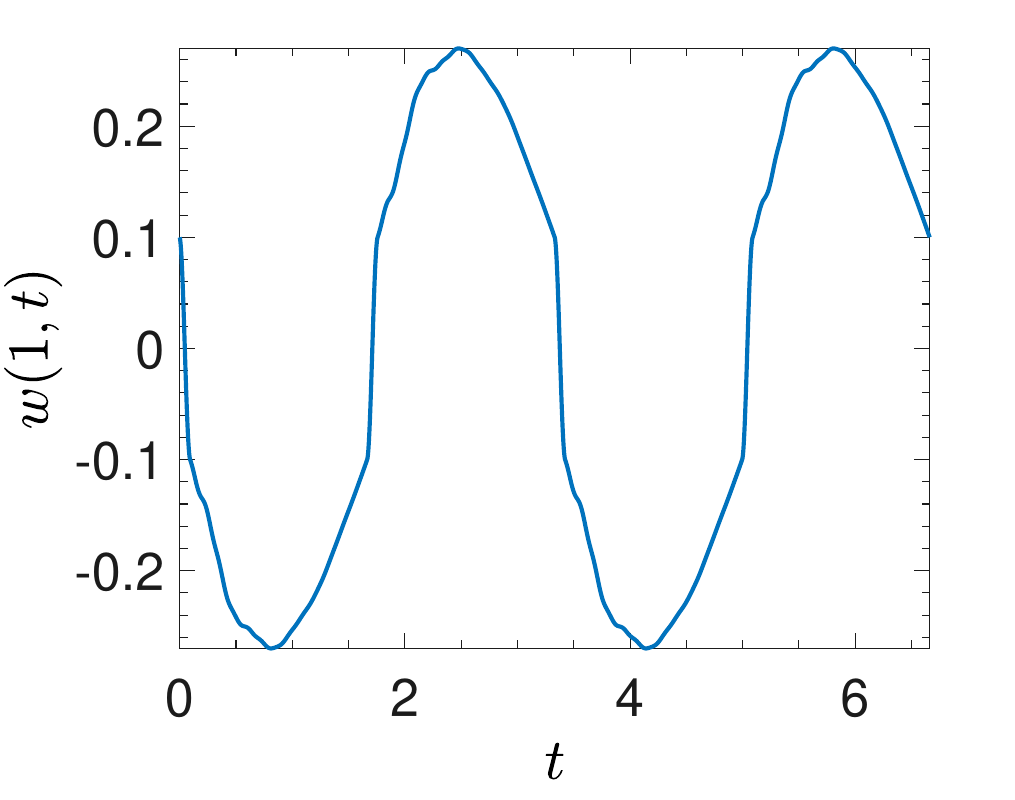}
	\caption{}
	\label{fig:F3a}
	\end{subfigure}
	\hfill
	\begin{subfigure}[b]{0.49\linewidth}
	\includegraphics[height=.8\linewidth]{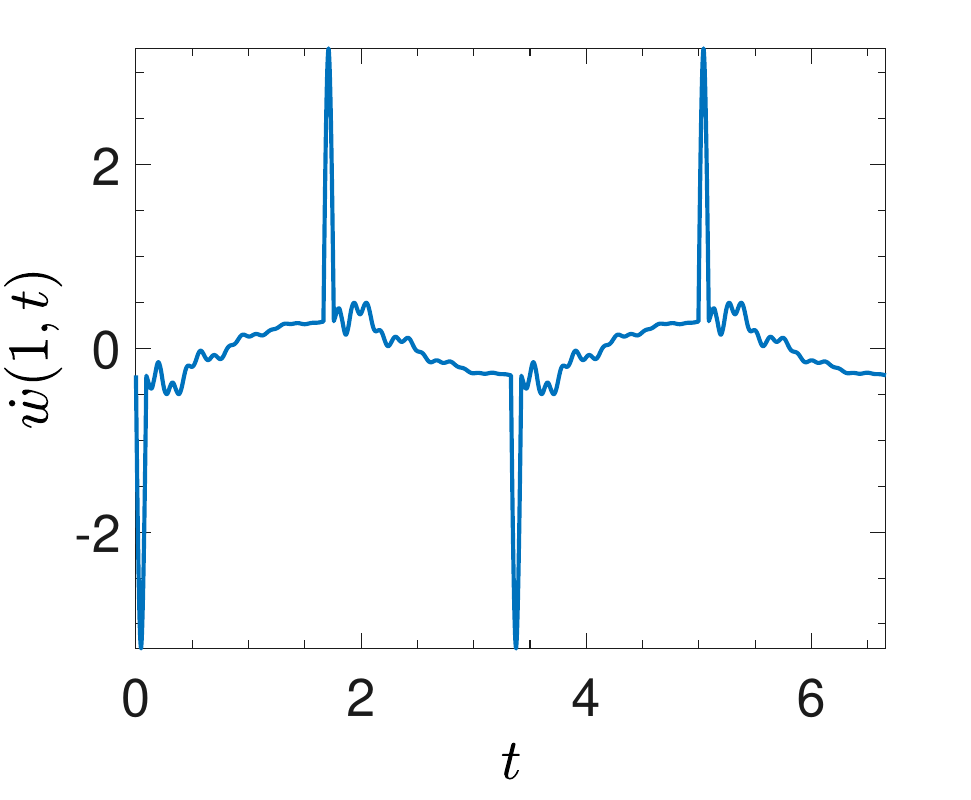}
	\caption{}
	\label{fig:F3b}
	\end{subfigure}
	\caption{Period-1, steady-state (a) displacement and (b) velocity time-series of the full order system, at the beam tip ($x=1$). The beam is subjected to a kick strength $F\approx 12.95$. The fundamental period of the time-series is $T\approx 3.25$ units. The remaining parameter values are same as in \cref{fig:bfd}.}
	\label{fig:FOS_3p}
\end{figure}

\begin{figure}[t!]
\centering
    \begin{subfigure}[t]{0.49\linewidth}
    \includegraphics[height=0.88\linewidth]{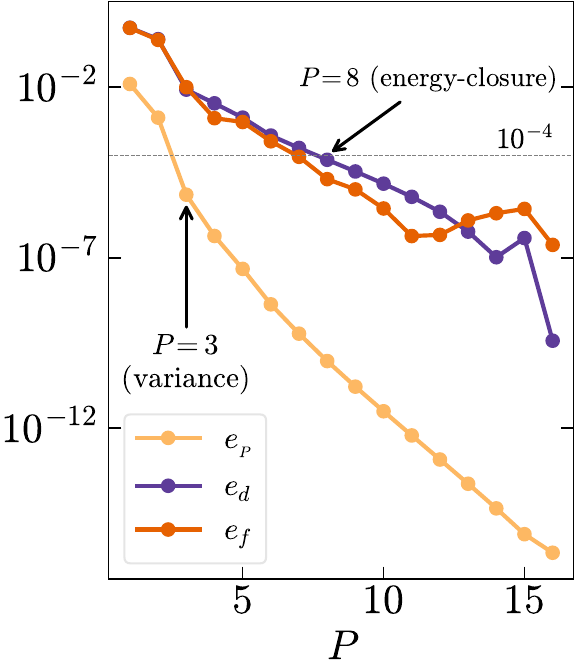}
    \caption{}
    \label{fig:edef}
    \end{subfigure}
    \begin{subfigure}[t]{0.49\linewidth}
    \includegraphics[height=0.89\linewidth]{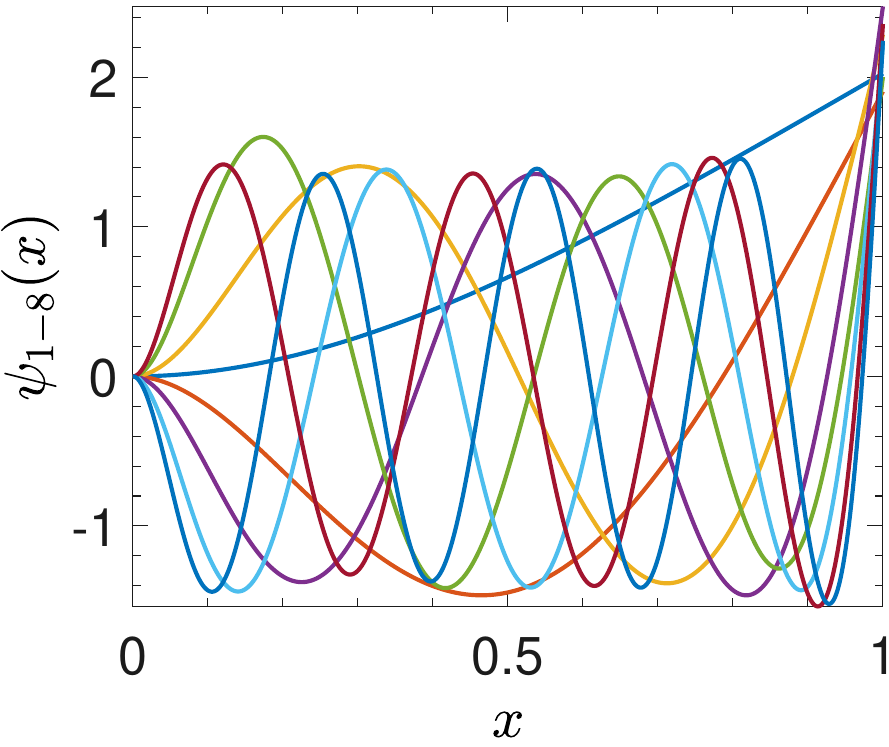}
    \caption{}
    \label{fig:poms}
    \end{subfigure}
    \caption{Energy closure analysis for the period-1 solution of \cref{fig:FOS_3p}: (a) The energy convergence errors ${e}_d$ and ${e}_f$ (Eq.~\ref{eq:edef}) vs. subspace dimension $P$. Both curves decrease below the tolerance $\epsilon_\text{tol} = 10^{-4}$ at $P=8$. In contrast, the variance error $e_{_P}$ (Eq.~\ref{eq:e_k}), shown only for comparison, decays much faster than both ${e}_d$ and ${e}_f$ and drops below the tolerance at $P=3$. (b) The eight POMs that span the subspace, obtained by applying POD on the displacement data.}
\end{figure}

Out of the range of steady-state responses shown \cref{fig:bfd}, we first selected the period-1 solution corresponding to $F\approx 12.95$ (see \cref{fig:pp_a}) to perform model reduction. 
Representative displacement and velocity time series are shown in \cref{fig:FOS_3p}. 
We selected the dimension of the ROM using the energy closure criterion: after initial experimentation, we selected $\epsilon_\text{tol} = 10^{-4}$; in \cref{fig:edef}, we plot ${e}_d$ and ${e}_f$ against $P$ and observe that both decrease in a monotonic fashion to this level. 
This indicates that $\widetilde{W}_d$ and $\widetilde{W}_f$ converge to their true values, $W_d$ and $W_f$, respectively, which further suggests that the subspaces have come into energy balance. 
For $P \ge 8 $, both ${e}_d$ and ${e}_f$ fall below the tolerance, indicating approximate energy closure, so we formulated an 8 DOF ROM following the steps described in the previous section. 

The corresponding 8 POMs (Fig.~\ref{fig:poms}) are \textit{global} empirical shape functions that span all regions of the system's hybrid phase space. However, they do not satisfy the force (natural) boundary conditions in any of the hybrid regions. Hence, our ROM basis consists of 8 admissible functions, in contrast with the total of 75 normal modes used in the ``exact'' simulations (25 for each set of boundary conditions). 
The fact that the POMs visually approximate the normal modes for a clamped-free beam, despite the more complex piecewise boundary conditions, likely indicates that the beam tip spends relatively little time interacting with the kicking and restoring forces localized near the static equilibrium position.

The maximum possible ROM dimension obtainable with from the displacement data set, $P_{\text{max}}=16$, indicated by the upper limit of the horizontal axis in \cref{fig:edef}, was determined from the rank of the displacement covariance matrix \cref{eq:disc_cov}. This means that, to within numerical precision, the displacement time series sampled at $N_\Omega = 100$ points along the beam contained information from \textit{at most} 16 statistically independent DOF. This confirms that $N=25$ normal modes was sufficient for convergence of the high-fidelity simulations and, at the same time, shows that an 8 DOF model provides a significant dimension reduction.

For the sake of comparison, we also employed the conventional variance-based mode selection criterion. 
The monotonically decaying $e_{_P}$ curve (displayed in yellow) in \cref{fig:edef} indicates that selecting only $P = 3$ modes is sufficient to capture more than $99.99\%$ ($e_{_P}<10^{-4}$) of the total response variance.
However, as seen in \cref{fig:edef}, when $P=3$, the energy convergence errors are at least two orders of magnitude greater than the tolerance level. 

\begin{figure}[tbp!]
\centering
	\begin{subfigure}[b]{0.48\linewidth}
	\includegraphics[height=0.85\linewidth]{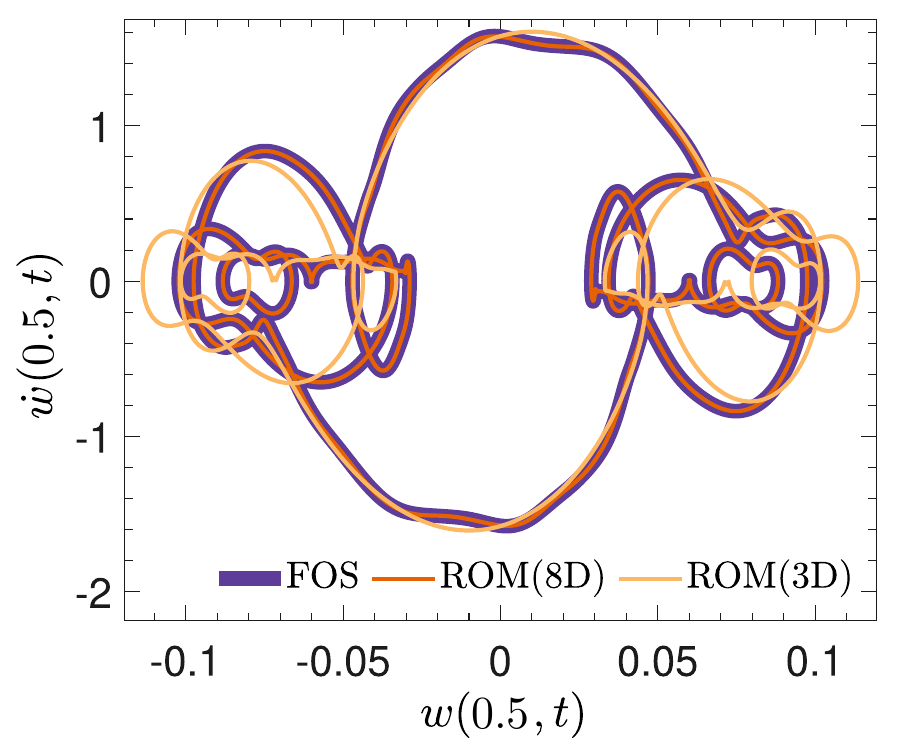}
	\caption{}
	\label{fig:PLS_ROM_Beam_mid_pp_8_3_fos}
	\end{subfigure}
	\hfill
	\begin{subfigure}[b]{0.48\linewidth}
	\includegraphics[height=0.85\linewidth]{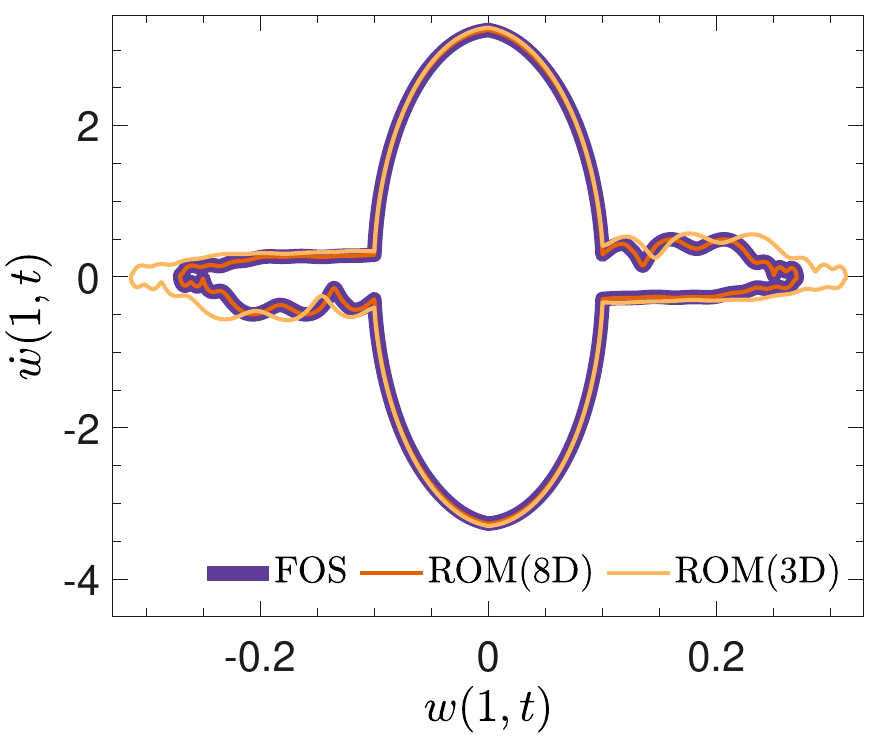}
	\caption{}
	\label{fig:PLS_ROM_Beam_tip_pp_8_3_fos}
	\end{subfigure}

	\caption{Phase plane plots illustrating the accuracy of reduced order models (ROMs) of the kicked oscillator, obtained by applying proper orthogonal decomposition (POD) to displacement data from period-1 solution of \cref{fig:pp_a,fig:FOS_3p}, with ROM dimension $P$ selected with either energy closure criterion or using 99.99\% of total variance. Energy closure gives $P=8$, while the variance-based criterion gives $P=3$. (a) Comparison of ROM responses to that of full order system (FOS) at $x=0.5$. (b) Same comparison at $x=1$. At both locations, the 3 DOF ROM gives a poor approximation to the response of the FOS (relative displacement and velocity field errors: $35.35\%$ and $114.71\%$, respectively); 8 DOF ROM response is much more accurate (relative displacement and velocity field errors: $0.002\%$ and $0.2\%$, respectively).}
\label{fig:pp_comparison}
\end{figure}

In \cref{fig:pp_comparison}, we use phase plane plots at the beam's midpoint and tip to contrast the dynamics of the full order system (FOS) against the 3 DOF ROM, formulated using a total variance criterion, and the 8 DOF ROM, formulated using the energy closure criterion (\cref{fig:PLS_ROM_Beam_mid_pp_8_3_fos,fig:PLS_ROM_Beam_tip_pp_8_3_fos}, respectively).
Upon inspecting both figures, it is clear that the 3 DOF ROM struggles to accurately reflect the period-1 dynamics exhibited by the FOS. The root mean square (RMS) errors for the displacement and velocity fields (that is, along the length of the beam) were computed for the 3 DOF ROM and found to be approximately $35.4\%$ and $114.7\%$, respectively. 
In contrast, the 8 DOF ROM mirrors the full order system with high fidelity, as reflected by the fact that the RMS errors for the displacement and velocity fields were $0.002\%$ and $0.2\%$, respectively. 
\begin{figure}[tbp!]
\centering
	\begin{subfigure}[b]{.9\linewidth}
	\centering
	\includegraphics[width=.9\linewidth]{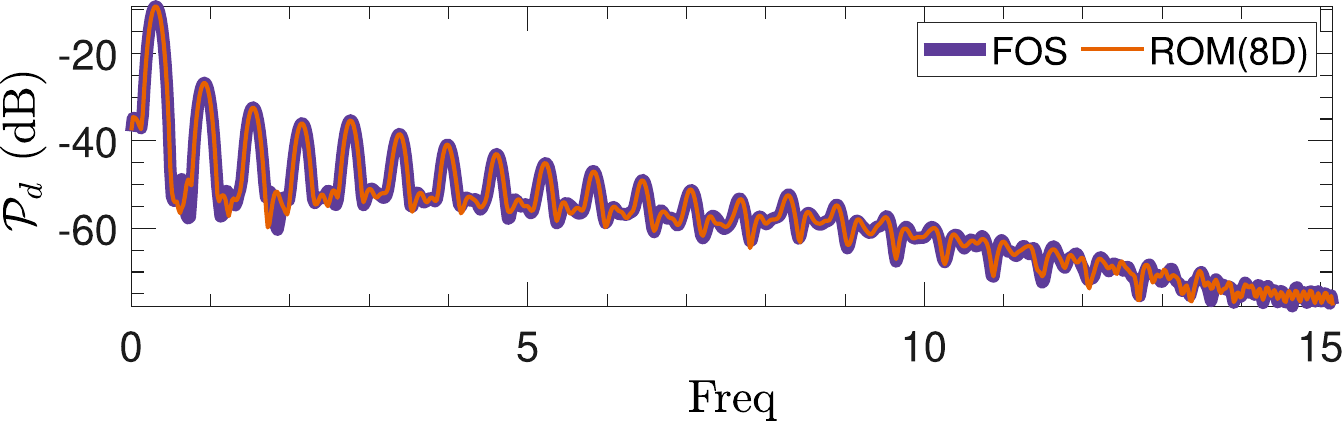}
	\caption{}
	\label{fig:rom_spec_da}
	\end{subfigure}
	\begin{subfigure}[b]{.9\linewidth}
	\centering
	\includegraphics[width=.9\linewidth]{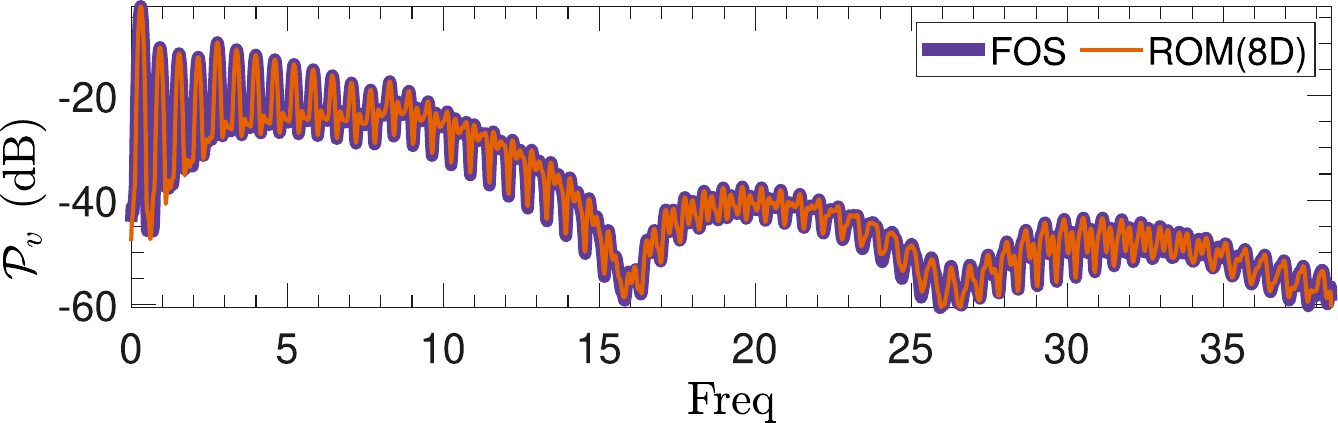}
	\caption{}
	\label{fig:rom_spec_va}
	\end{subfigure}
	\begin{subfigure}[b]{.9\linewidth}
	\centering
	\includegraphics[width=.9\linewidth]{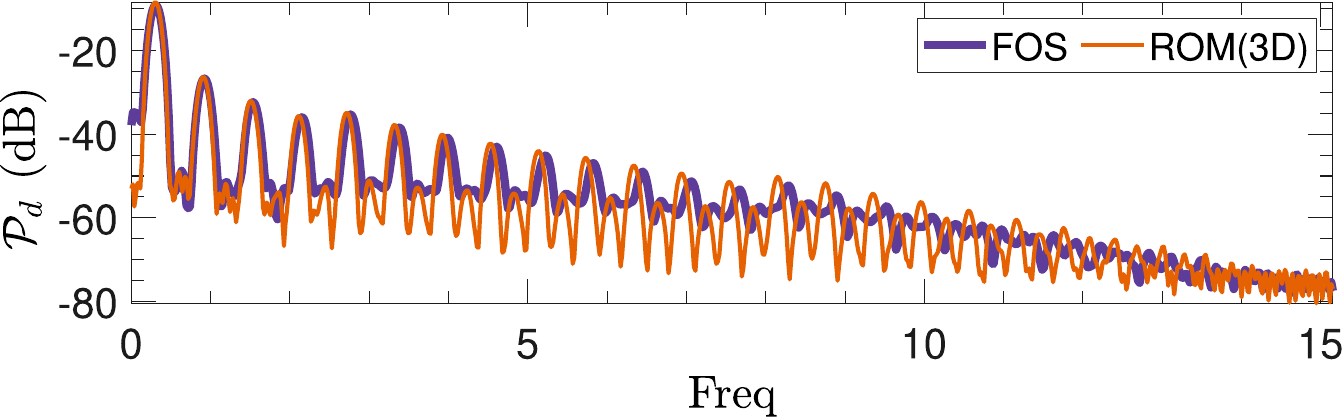}
	\caption{}
	\label{fig:rom_spec_db}
	\end{subfigure}
	\begin{subfigure}[b]{.9\linewidth}
	\centering
	\includegraphics[width=.9\linewidth]{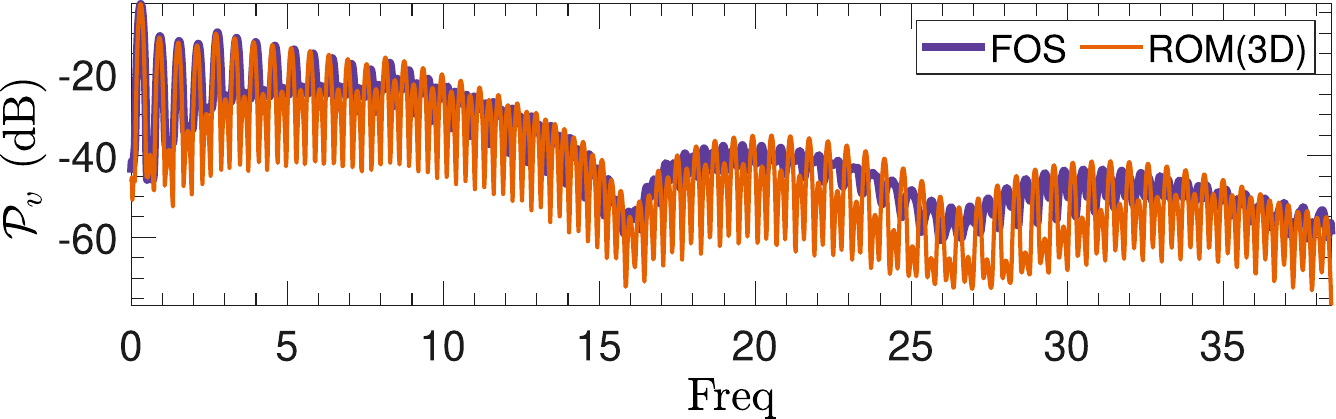}
	\caption{}
	\label{fig:rom_spec_vb}
	\end{subfigure}
	
	\caption{Comparison of the power spectra for the full order system (FOS) and 8 DOF ROM: (a) displacement and (b) velocity at the beam tip. Power spectra comparison between FOS and 3 DOF ROM: (c) displacement and (d) velocity time-series at the beam tip. Time series are from the solutions of \cref{fig:PLS_ROM_Beam_tip_pp_8_3_fos}.}
\label{fig:ROM_psd}
\end{figure}

The same outcome is observed while comparing power spectra of the displacement and velocity beam tip time series, as shown in \cref{fig:ROM_psd}. The fundamental frequency of the response of the FOS, shown in purple, is approximately $0.307$, corresponding to a period $T\approx 3.25$, as can also be seen in the time series plots of  \cref{fig:FOS_3p}. The other peaks are at odd harmonics of the fundamental frequency. 
From \cref{fig:rom_spec_da,fig:rom_spec_va} we see the frequency content of the response obtained from the 8 DOF ROM is nearly identical to the FOS: all peaks in the frequency domain are almost perfectly aligned with those of the FOS as shown by the red line, indicating high accuracy of the ROM in capturing different time scales of the FOS response. 
In contrast, if we compare the frequency content of the response from the 3 DOF ROM, as shown in \cref{fig:rom_spec_db,fig:rom_spec_vb}, we observe that the overall structure of the power spectrum is significantly different and even though the first five or six frequency peaks align with those of the FOS, the peaks at higher frequencies diverge substantially.
Furthermore, even the low-amplitude frequency content between peaks is accurately captured by the 8-DOF model; these are very accurate for the 3 DOF ROM determine by variance only.
Thus, the 8 DOF ROM formulated using energy closure clearly outperforms the 3 DOF ROM using 99.99\% of the total variance. 
Now, obviously, in practise one can try increasing the fraction of total variance to formulate a ROM with comparable accuracy; however, the point here is that energy closure provides a physically grounded explanation as to \textit{why} the additional modes are required.

\begin{figure}[t!]
	\centering
	\includegraphics[width=1\textwidth]{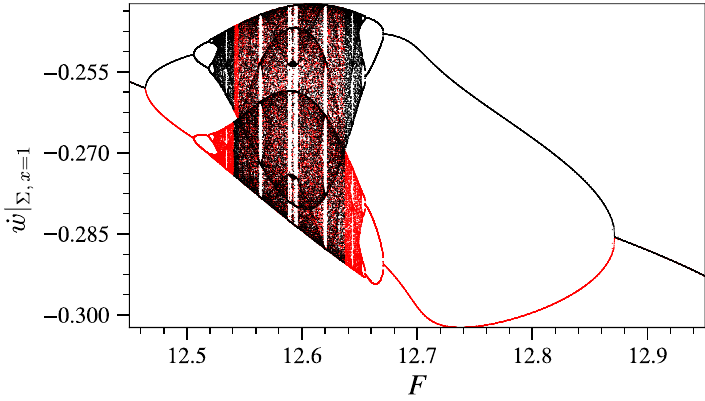}
	
	\caption{Reconstructed bifurcation diagram of the kicked oscillator (\cref{fig:bfd}), in this case using simulations of an 8 DOF reduced order model (ROM) for kick strengths $F\in[12.44,12.95]$. The ROM was constructed with proper orthogonal decomposition (POD) and energy closure using only displacement data from the period-1 solution at $F = 12.95$ (\cref{fig:pp_a}). Despite this limited data set, the reconstructed diagram is virtually identical to the bifurcation diagram of the full order system (FOS).}
	\label{fig:rom_fig:bfd}
\end{figure}

A major open question when creating ROMs in nonlinear systems is how robust they are with respect to changes in parameters, particularly across bifurcations. 
We examined if the 8 DOF ROM formulated above, which is based only on data from one period-1 solution for $F=12.95$, can accurately approximate other steady-state responses of the FOS corresponding for other values of $F$. 
To this end, we attempted to reconstruct the bifurcation diagram of \cref{fig:bfd} using the 8 DOF ROM. 
The result is shown in \cref{fig:rom_fig:bfd}: we see that the 8 DOF ROM is capable of recovering the complete bifurcation structure found using the full order system (FOS).

To further study the robustness of our model reduction procedure, we used POD and energy closure analysis to construct different subspaces for each of each of the $10^4$ solutions represented in \cref{fig:bfd}. Using the same energy closure tolerance, $\epsilon_\text{tol}=10^{-4}$, all subspace dimensions were found to be $P=8$. We further checked the angles between each of these 8-dimensional subspaces: in all cases, these were found to be negligible, on the order of $10^{-3}$ radians. 
This implies that these subspaces essentially defined the same hyperplane, to within numerical error, further implying that the different sets of POMs spanning these subspaces were at most rotations of each other. These results help explain why the original 8 DOF ROM formulated using $F=12.95$ data was capable of recovering the full bifurcation diagram for the system. Furthermore, they strongly suggest that ROMs formulated using data from any of point in the bifurcation diagram would be able to do the same. Thus, we found that the reduced order modeling procedure is highly robust to changes in the kick strength $F$.

This degree of robustness was at first surprising to the authors; however, a similar phenomenon has been observed in studies of the bifurcation structure of the complex Ginzburg-Landau equation \cite{terragni_efficient_2014,Rapun2011,Terragni2012}. 
In these studies, it was found that POMs calculated for specific values of the bifurcation parameter could be used to simulate the system with other parameter values, as long as a sufficient number of POMs were retained. Furthermore, the researchers found that the corresponding ROMs could be expected to maintain their accuracy in some neighborhood of bifurcation points, and were able to approximate  different attractors of the system. In our study, we observe that it is energy closure that identifies this ``sufficient number'' of POMs capable of capturing  the dynamics in the \textit{entire} parameter range of interest. The robustness of the ROMs observed in our solid mechanical system is an important addition to the previously reported instances in the context of fluid systems. 

Thus, our results suggest that a robust construction of ROMs requires the inclusion of a sufficient number of DOF across the entire parameter range of interest and, furthermore, that these DOF should lie in the same subspace. Our study shows that energy closure analysis provides a physics-based criterion capable of doing precisely this.

\section{Summary and conclusions}

We investigated the dynamics of a kicked, piecewise linear, globally nonlinear flexible oscillator and its model reduction using proper orthogonal decomposition (POD), with energy closure analysis used for dimension estimation \cite{Bhattacharyya_2020,Bhattacharyya_2022}.
Our first principles model consists of a clamped-free Bernoulli-Euler beam with a point mass at its tip, subjected to a position-triggered ``kick'' of adjustable strength and a localized linear restoring force acting near the beam's static equilibrium. 
The resulting nonlinear infinite-dimensional hybrid system is defined by how the piecewise tip forces create 8 regions in phase space (Fig.\ref{fig:phse}), described by 3 distinct linear models with different boundary conditions (Eqs.\ref{eq:weak_eom} and \ref{eq:weak_eom_cases}).

We represented the full-order system dynamics with high-fidelity numerical simulations of 25-degree-of-freedom (DOF) models constructed with the linear normal modes defined in each region of the hybrid state space. The full range of steady state behaviors of the system was obtained and summarized with a bifurcation diagram using the kick strength $F$ as the bifurcation parameter (Fig.~\ref{fig:bfd}), revealing a complex structure of stability transitions between various periodic and chaotic steady states.

Displacement field data from a period-1 steady state at $F=12.95$ was used to construct global empirical basis functions via POD: an 8-DOF model was found to satisfy energy closure to within a specified tolerance. These ROMs accurately captured the full-order system dynamics at the chosen steady state while achieving an approximately 300\% computational speedup. More significantly, even though data from only a single periodic steady state was used in its construction, the ROM was found to reproduce the full bifurcation structure (Fig.~\ref{fig:rom_fig:bfd}). In addition, POD with a fixed energy closure tolerance was found to select subspaces of the same dimension spanning nearly identical hyperplanes across the bifurcation parameter range, indicating strong uniformity in the   system's effective degrees of freedom.

In contrast, a 3 DOF ROM estimated using a conventional 99.99\% variance criterion failed to replicate the true response or the bifurcation structure. While raising the retained variance (e.g., to 99.999\%) can increase DOFs, such a statistical approach is inferior to energy closure in at least two respects. First, energy closure provides a physics-based rationale for selecting a specific number of empirical modes: a low-variance mode may still play a critical energetic role. Second, different steady states in the same system may require different variance levels whereas, at least for this system, the same energy closure tolerance produced accurate ROMs and recovered the bifurcation structure. This suggests energy closure may offer a uniform bound on ROM performance in a way variance does not.

As this work is purely analytical and numerical, we used full-state field data. However, even with a first-principles model, noisy or insufficient experimental data may hinder accurate energy closure estimation. In \cite{Bhattacharyya_2022}, we addressed this by outlining how to proceed with limited data (e.g., missing velocities), strategies to reduce measurement noise, and efficient estimation methods for noisy inputs.

In this paper, we leveraged periodicity to simplify computations: energy input and dissipation (Eq.~\ref{eq:WdWf}) were evaluated using data from a single steady-state period. However, the assessment of energy closure is not limited to periodic data: in principle, it can be determined for chaotic or stochastic behaviors, though this generally will require more data to estimate the average energy balance accurately. Indeed, for the $10^4$ values of $F$ in Fig.~\ref{fig:bfd}, energy closure for chaotic responses was evaluated using data from all 16 forcing periods used in constructing the bifurcation diagram, which was sufficient for statistical convergence. However, for deterministic systems, a key outcome of the ROMs’ robustness to parameter variations is that only periodic responses need be analyzed, keeping data collection requirements to a minimum.

A fundamental limitation of our approach is that only steady state data was used to construct ROMs and, thus, we cannot expect our models to reliably model arbitrary transient behavior. However, this is a limitation of any model reduction procedure because, in general, it will always be possible to start transients from unmodeled degrees of freedom far away from attracting steady states. On the other hand, the fact that our ROMs reproduce the entire bifurcation structure indicates that they \textit{are} capable of capturing transient dynamics starting sufficiently close to steady states: if this were not true, at each stability transition the ROM would have not approached the correct attractor, as needed to generate the bifurcation diagram using simple continuation. 

\bibliographystyle{unsrtnat}
\bibliography{refs,refs_MOR2,refs_PW_oscillators2,refs_revision}

\section*{Appendices}
\appendix

\section{Nondimensionalization of governing equations}
\label{sec:AppC_ND_derv}
Implementing Hamilton's principle, the governing partial differential equation (PDE) of a cantilever Bernoulli-Euler beam of length $l$, with point mass $m$ and spring to ground of stiffness $k$, both at the tip, as shown in \cref{fig:es_b}, can be written in \textit{dimensioned} form as:
\begin{equation}
	\rho A\ddot{w}+EIw''''+c_{m}EI\dot{w}''''+c_{v}\dot{w}=0,
	\label{eq:AppC_dim_eom}
\end{equation}
where $w(x,t)$ denotes the horizontal deflection of the beam during oscillation measured from the vertical neutral axis,  with $x\in[0,l]$; primes and overdots indicate derivatives with respect to $x$ and $t$, respectively; $c_m$ and $c_v$ denote coefficients of material damping and viscous damping, respectively; $E$ is Young's modulus; $I$ is the cross-sectional area moment of inertia of the beam; $A$ is area of cross section; and $\rho$ is the mass density of beam material. 

Hamilton's principle also provides dimensioned boundary conditions. As discussed in Sec.~\ref{sec:pw}, and summarized in Fig.~\ref{fig:phse}, the system has a hybrid set of boundary conditions. For the purposes here, we need only focus on model B, corresponding to the system when the kick force is active (Eqs.~\ref{eq:pw_bcs}):
\begin{align}
	w(0,t) & =0\nonumber \\
	w'(0,t) & =0\nonumber \\
	EI\left( w''(l,t)+c_{m}\dot{w}''(l,t)\right)  & =0\nonumber \\
	EI\left( w'''(l,t)+c_{m}\dot{w}'''(l,t)\right)  & =m\ddot{w}(l,t)+kw(l,t)+\text{sgn}({\dot{w}(l,t)})F,\label{eq:AppC_dim_BCs}
\end{align}
To non-dimensionalize the above system, we indicate dimensionless variables by an overbar and define $x=l\,\overline{x}$, $w=l\, \overline{w}$ and $t=t_{c}\,\overline{t},$ where $t_c$ is to be found. The differential operators then become
\[
\frac{\partial}{\partial x}=\frac{1}{l}\frac{\partial}{\partial\,\overline{x}}\quad\text{and}\quad\frac{\partial}{\partial t}=\frac{1}{t_{c}}\frac{\partial}{\partial\, \overline{t}}\,,
\]
so \cref{eq:AppC_dim_eom} can be written as 
\begin{align*}
	\rho A\frac{\ddot{\overline{w}}}{t_{c}^{2}}+EI\frac{\overline{w}\,''''}{l^{4}}+c_{m}EI\frac{\dot{\overline{w}}\,''''}{l^{4}t_{c}}+c_{v}\frac{\dot{\overline{w}}}{t_{c}} & =0\\
	\implies \ddot{\overline{w}}+t_{c}^{2}\frac{EI}{\rho Al^{4}}\overline{w}\,''''+c_{m}t_{c}\frac{EI}{\rho Al^{4}}\dot{\overline{w}}\,''''+\frac{c_{v}t_{c}}{\rho A}\dot{\overline{w}} & =0
\end{align*}
in which and overdots and primes now represent \textit{dimensionless} derivatives. Letting  $\overline{c}_{m}=c_m/t_c$, $\overline{c}_{v}=c_{v}\mspace{2mu}t_c /(\rho A)$ and $t_{c}=\sqrt{\rho A l^{4}/(EI)}$, the PDE can be written in \textit{dimensionless} form as
\begin{equation}
	\ddot{\overline{w}}+\overline{w}\,''''+\overline{c}_{m}\dot{\overline{w}}\,''''+\overline{c}_{v}\dot{\overline{w}}=0.
	\label{eq:ND_EOM}
\end{equation} 
Correspondingly, the BCs take the dimensionless form
\begin{align}
	\overline{w}(0,t) & =0\nonumber \\
	\overline{w}\,'(0,t) & =0\nonumber \\
	\overline{w}\,''(1,\overline{t})+\overline{c}_{m}\dot{\overline{w}}\,''(1,\overline{t})  & =0\nonumber \\
	\overline{w}\,'''(1,\overline{t})+\overline{c}_{m}\dot{\overline{w}}\,'''(1,\overline{t})  & = \overline{m}\,\ddot{\overline{w}}(1,\overline{t})+\overline{k}\overline{w} (1,\overline{t}) + \text{sgn}(\dot{\overline{w}}(1,\overline{t})\overline{F},\label{eq:Bc1}
\end{align}
in which we have also rescaled the mass, spring stiffness, and kick force using $\overline{m} = m/(\rho A l)$, $\overline{k} = k l^3/(EI)$, and $\overline{F} =  F l^2/(EI)$, respectively.

After rescaling, for convenience, we \textit{drop overbars}, which yields the dimensionless PDE and boundary conditions, \cref{eq:goveq,eq:pw_bcs}, presented in Sec.~\ref{sec:md}.\

\section{Removal of force boundary condition in kicking region}
\label{sec:asimub}
As discussed in Sec.~\ref{sec:pw}, when finding the solution for model B, the force boundary condition (BC) in the kicking region can be easily removed by expressing the displacement field $w(x,t)$ relative to the the static deflection associated with the constant tip force $F$. 

Depending on the sign of the velocity, the static deflection in this region is given by the time-independent version of  \cref{eq:goveq,eq:pw_bcs}:
\[
w''''(x)=0
\]
and
\begin{equation}
\begin{aligned}
	{w}(0) & =0 \\
	{w}'(0) & =0 \\
	{w}''(1)  & =0\\
	{w}'''(1) & =k{w}(1)\pm F,
\end{aligned}
\end{equation}
in which the sign in front of $F$ is to be chosen depending on the sign of the tip-velocity. We write the solution to the above boundary value problem as
\begin{equation}
	w_s(x)=\pm\frac{3F}{3+k}\left( \frac{x^2}{2}-\frac{x^3}{6} \right).
\end{equation}
Then, defining the relative displacement $u(x,t)$ with respect to $w_s(x)$ via $w(x,t)=w_s(x)+u(x,t)$, substitution into the  second BC of \cref{eq:pw_bcs_cases} (corresponding to regions 4, 8) results in a BC in $u$ identical in form to that of the third BC of \cref{eq:pw_bcs_cases} (corresponding to regions 1, 3, 5, 7), as was to be shown.

\section{Orthonormality conditions}
\label{sec:AppC1}
For the modal analysis used in the piecewise solution of Sec.~\ref{sec:pw}, we employed orthonormality conditions Eqs.~\eqref{eq:normalize_ms}: we here summarize their derivation. The relevant inner products are obtained by looking at conservative free vibrations of the dimensionless system (Sec.~\ref{sec:AppC_ND_derv}) of \cref{eq:goveq,eq:pw_bcs}, that is, the PDE
\begin{equation}
	w''''(x,t)+\ddot{w}(x,t)=0,\label{eq:ip1}
\end{equation}
subjected to the boundary conditions (BCs)
\begin{equation}
\begin{aligned}
	{w}(0,t) & =0 \\
	{w}'(0,t) & =0 \\
	{w}''(1,\bar{t})  & =0\\
	{w}'''(1,\bar{t}) & =m\ddot{{w}}(1,{t})+k{w}(1,{t}).
\end{aligned}
\end{equation}
Assuming a time-harmonic response, separation of variables $w(x,t)=W(x)e^{i \omega t}$ yields the differential eigenvalue problem 
\begin{equation}
	W''''(x)-\omega^{2}W(x)=0,\label{eq:ip2}
\end{equation}
with BCs
\begin{equation}\label{eq:BVP_BCs}
\begin{aligned}
	W(0) & =0 \\
	W'(0) & =0 \\
	W''(1)  & =0\\
	W'''(1) & = - m \omega^2 W(1)+k W (1),
\end{aligned}
\end{equation}
the solutions to which yield the natural frequencies $\omega_i$ and normal modes $W_i$ ($i=1,2,3,\ldots$). Substituting the $i^\text{th}$ natural frequency and normal mode in \cref{eq:ip2}, multiplying by the $j^\text{th}$ normal mode, and integrating over the spatial domain of the problem (the length of the beam), we get
\begin{equation}
	\int_{0}^{1}W_{j}W_{i}''''dx-\omega_{i}^{2}\int_{0}^{1}W_{j}W_{i}dx=0.\label{eq:ip4}
\end{equation}
Integrating by parts two times and using the BCs (Eq.~\ref{eq:BVP_BCs}), we obtain
\begin{equation}
	\int_{0}^{1}W_{i}''W_{j}''dx + k W_{i}(1)W_{j}(1) = \omega_{i}^{2} \left [\, \int_{0}^{1}W_{i}W_{j}dx + m W_{i}(1)W_{j}(1) \,\right ]. \label{eq:ip5}
\end{equation}
Given that in this case $\omega_i \ne \omega_j$ for $i \ne j$, swapping the indices $i$ and $j$ and subtracting the result from the above yields the orthogonality condition satisfied by the normal modes:
\begin{equation}\label{eq:KE_innerprod}
\int_{0}^{1}W_{i}W_{j}dx+mW_{i}(1)W_{j}(1)=\delta_{ij}.
\end{equation}
Substitution of the above into \cref{eq:ip5} then yields the additional condition:
\begin{equation}\label{eq:PE_innerprod}
\int_{0}^{1}W_{i}''W_{j}''dx+k W_{i}(1)W_{j}(1)=\omega_i^2 \delta_{ij}.
\end{equation}
It can be immediately observed that, subject to the appropriate assumptions on the stiffness $k$, \cref{eq:KE_innerprod,eq:PE_innerprod} have the same form as  Eqs.~\eqref{eq:normalize_ms}.

\end{document}